\documentclass[]{emulateapj}
  \usepackage{rotating}
\newcommand{\feh}{\mbox{[Fe/H]}}
\newcommand{\zh}{\mbox{[Z/H]}}
\newcommand{\afe}{\mbox{[$\alpha$/Fe]}}

\slugcomment{Accepted for publication in ApJ}

\shorttitle{VLT Spectroscopy of Globular Clusters in Low-Surface Brightness Dwarf Galaxies} 
\shortauthors{Puzia \& Sharina}

\begin{document}
\title{VLT Spectroscopy of Globular Clusters in Low Surface Brightness Dwarf 
Galaxies\altaffilmark{\mbox{$\star$}}}
\altaffiltext{$\star$}{Based on observations made with ESO Telescopes at the Paranal Observatory under program ID P76.AB-0137.}
\author{Thomas H. Puzia\altaffilmark{1,2} \& Margarita E. Sharina\altaffilmark{3,4}}
\altaffiltext{1}{Herzberg Institute of Astrophysics, 5071 West Saanich Road, Victoria, BC V9E 2E7, Canada}
\altaffiltext{2}{Plaskett Fellow}
\altaffiltext{3}{Special Astrophysical Observatory, Russian Academy of Sciences, N.Arkhyz, KChR, 369167, Russia}
\altaffiltext{4}{Isaac Newton Institute of Chile, SAO Branch}

\begin{abstract}
We present VLT/FORS2 spectroscopic observations of globular clusters (GCs)
in five low surface brightness (LSB) dwarf galaxies: KK211 and KK221,
which are both dwarf spheroidal satellites (dSph) of NGC~5128, dSph KK84
located close to the isolated S0 galaxy NGC~3115, and two isolated dwarf
irregular (dIrr) galaxies UGC~3755 and ESO~490-17.~Our sample is selected
from the Sharina et al. (2005) database of Hubble Space Telescope WFPC2
photometry of GC candidates in dwarf galaxies. For objects with accurate
radial velocity measurements we confirm 26 as genuine GCs out of the 27
selected candidates from our WFPC2 survey.~One candidate appears to be a
distant galaxy.~Our measurements of the Lick absorption line indices in
the spectra of confirmed GCs and the subsequent comparison with SSP model
predictions show that all confirmed GCs in dSphs are old, except GC
KK211-3-149 ($6 \pm$2 Gyr), which we consider to be the nucleus of KK211.
GCs in UGC~3755 and ESO~490-17 show a large spread in ages ranging from
old objects ($t>10$ Gyr) to clusters with ages around 1 Gyr. Most of our
sample GCs have low metallicities $\zh \le -1$. Two relatively metal-rich
clusters with $\zh \approx -0.3$ are likely to be associated with
NGC~3115. Our sample GCs show in general a complex distribution of
$\alpha$-element enhancement with a mean
$\langle$[$\alpha$/Fe]$\rangle=0.19\pm0.04$ derived with the $\chi^{2}$
minimization technique and $0.18\pm0.12$ dex computed with the iterative
approach. These values are slightly lower than the mean
$\langle$[$\alpha$/Fe]$\rangle=0.29\pm0.01$ for typical Milky Way GCs. We
compare other abundance ratios with those of Local Group GCs and find
indications for systematic differences in N and Ca abundance. The specific
frequencies, $S_N$, of our sample galaxies are in line with the
predictions of a simple mass-loss model for dwarf galaxies and compare
well with $S_N$ values found for dwarf galaxies in nearby galaxy clusters.
\end{abstract}

\section{Introduction}
\label{intro}
The hierarchical structure formation scenario predicts that dwarf galaxies
are the first systems to form in the Universe \citep{peebles68}, and that
more massive galaxies form through dissipative processes from these
smaller sub-units. The involved physical mechanisms of this sequence
depend on the density and mass of the parent dark matter halo, in the
sense that more massive halos initiate star formation at earlier epochs
and form their stars at a faster rate \citep[e.g.][]{peebles02, renzini06,
ellis07}. Because of this environmental gradient, we expect that dwarf
galaxies in the field formed the first stellar population relatively late
and at a lower pace compared to their counterparts in dense galaxy
clusters. In other words, the difference in age and chemical composition
between the oldest stellar populations in cluster and field dwarf galaxies
should reflect the delay in the onset of structure formation in these two
environments.

The task of measuring the age and chemical composition of the oldest
stellar populations in distant dwarf galaxies from their integrated light
is very challenging. An alternative approach is to investigate the oldest
globular clusters (GCs) that are found in dwarf galaxies. Several
photometric surveys of extragalactic GCs in dwarf galaxies outside the
Local Group have been performed in the past decade \citep[see review
by][]{miller06}. However, only a handful of those were followed up with
8--10m-class telescopes to derive spectroscopic ages and chemical
composition. Observations of galaxies in groups and clusters provide more
and more evidence that environment is a major factor influencing the
process of GC formation \citep[e.g.][]{west93, tully02, grebel03,
miller98}.~Recent progress in modeling the assembly history of GC systems
in massive elliptical galaxies suggests that a significant fraction of
metal-poor GCs were accreted from dwarf satellites at later times compared
to the number of GCs initially formed in the parent galaxy halo
\citep{ppm07}. These results underline the ideas put forward in the work
of \cite{forte82} and \cite{muzzio87} as well as the models of
\cite{cote98, cote02, hilker99}, who suggested that the rich GC systems of
massive galaxies may be the result of significant GC accretion through
tidal stripping of less massive systems.

Spectroscopic studies of a few GCs in cluster and field dwarf galaxies
showed that most of these systems host at least some old GCs with ages
$t\ga10$ Gyr \citep{puzia00, sharina03, strader03a, strader03b, beasley06,
conselice06}.~Although today's accuracy of relative spectroscopic age
determinations ($\Delta t/t \approx 0.2-0.3$) is not sufficient to resolve
the expected delay of $\sim\!0.5-4$ Gyr in the onset of star-formation
between cluster and field environment \citep[depending on cosmology,
ionizing source population, ionization feedback efficiency, etc.,
see][]{kauffmann96, treu05, thomas05, delucia06, clemens06}, the old ages
combined with information on abundance ratios can provide a powerful tool
to decide whether stellar populations in field dwarf galaxies followed the
same early enrichment history as their analogs in denser environments.
Furthermore, any difference in GC chemical composition between dwarf and
more massive galaxies opens an attractive way of chemically tagging
accreted sub-populations in massive halos and, therefore, enables us to
quantify the mass accretion history of galaxies, a task that for old
galaxies is infeasible from studies of the diffuse galaxy light alone.
Similar ideas have been formulated for stellar populations that make up
the diffuse component of the most nearby galaxies, which are close enough
for high-resolution spectroscopy of individual stars \citep{fh02,
geisler07}. The obvious advantage of GC systems is that their spectra can
be observed out to about 10 times greater distances.

In this work we analyze spectroscopic observations of GCs in dwarf
irregular (dIrr) and dwarf spheroidal galaxies (dSph) in the field/group
environment. Our sample consists of systems that are representatives of
the lowest-mass bin of the Local Volume (LV) galaxy population, limited to
distances $D<10$ Mpc. We refer to \cite{karachentseva85} and
\cite{grebel99} for a morphological type definition of our sample
galaxies. The paper is organised as follows. In Section~\ref{observations}
we describe our observations and data reduction as well as the methods of
measuring radial velocities. Section~\ref{analysis} summarizes the
measurement of Lick line indices, their calibration, and the determination
of spectroscopic ages, metallicities, and abundance ratios.
Section~\ref{discussion} is devoted to the discussion of our results.
Conclusions are presented in section~\ref{conclusion}.

\begin{deluxetable*}{lccccccclcr}[!ht]
\tabletypesize{\scriptsize}
\tablecaption{Properties of sample dwarf galaxies \label{dwgprop}}
\tablewidth{0pt}
\tablehead{
\colhead{Galaxy} & \colhead{RA (J2000)}& \colhead{DEC (J2000)} & 
\colhead{$ \mu_0$} & \colhead{$D_{\rm MD}$} &
\colhead{$A_{B}$} & \colhead{$B$} & \colhead{$B-I$} &
\colhead{$M_V$} & \colhead{$N_{\rm GC}$} & \colhead{$S_{N}$}
}
\startdata
KK211, AM1339-445   & 13 42 06 & $-$45 13 18& 27.77     & 0.25       & 0.477   & 16.3$\pm$0.2 & 1.8$\pm$0.2 & $-$12.58&  2  & 18.6\\
KK221               & 13 48 46 & $-$46 59 49& 28.00     & 0.50       & 0.596   & 17.3$\pm$0.4 & 2.0$\pm$0.4 & $-$11.96&  6  & 95.1\\
KK084, UA200, KDG65 & 10 05 34 & $-$07 44 57& 29.93     & 0.03       & 0.205   & 16.4$\pm$0.2 & 1.7$\pm$0.2 & $-$14.40&  7  & 10.4\\
UGC3755, PGC020445  & 07 13 52 & $+$10 31 19& 29.35     &$\sim5$     & 0.384   & 14.1$\pm$0.2 & 1.1$\pm$0.2 & $-$16.12&  32 & 11.4\\\smallskip
E490-017, PGC019337 & 06 37 57 & $-$25 59 59& 28.13     & 4.50       & 0.377   & 14.1$\pm$0.2 & 1.1$\pm$0.2 & $-$14.90&  5  &  5.4 \\
\enddata
\tablecomments{Columns contain the following data: (1) galaxy name, (2) equatorial
coordinates, (3) and (4) are the distance modulus and distance from the
nearest bright galaxy in Mpc from \cite{kara04}, and from \cite{tully05}
for UGC3755, (5) reddening from \cite{sch98} in the B-band, (6) is
the integrated $B$ magnitude, and (7) the integrated $B-I$ color, derived
from surface photometry on the VLT-FORS2 images in this work (see
Appendix~\ref{sbprof}), (8) absolute $V$ magnitude from SPM05, (9) number
of GCs according to SPM05 and this paper, (10) specific frequency,
$S_N=N_{\rm GC}10^{0.4(M_V+15)}$ \citep{harris81}.}
\end{deluxetable*}

\section{Observations}
\label{observations}

\subsection{Sample Selection}
\label{obs_data_red}
All target galaxies are part of the \cite{sh05} (hereafter: SPM05) sample
for which detailed information on luminosities, colors, and structural
parameters of GC candidates is available from HST data. The targets were
selected based on the number of GC candidates and the optimization of the
observing strategy at the time of scheduled observations. The main
characteristics of our sample galaxies are presented in
Table~\ref{dwgprop}. KK211 and KK221 are two of the faintest dSph galaxies
in the Centaurus~A group. The dSph galaxy KK84 is the nearest satellite of
the giant field S0 galaxy NGC~3115. UGC~3755\footnote{Strictly speaking
UGC~3755 is a borderline low surface brightness galaxy according to the
definition outlined by \cite{impey97}:
$\bar{\mu}_{0,B} \ge 22.5 - 23$ mag/arcsec$^{2}$.}, which hosts one of the
richest GC system among isolated dwarf galaxies (see Table~\ref{dwgprop}),
and ESO490-17 are both highly isolated dIrrs. In general, the specific
frequencies are high for all our sample galaxies. It will be shown in the
last section, that these $S_N$ values correspond to the predictions of
galaxy evolution models that include significant mass loss, which strongly
affects the star formation processes in low-mass galaxies.

Detailed CMD studies of the Local Group dwarf galaxies show that each
low-mass galaxy has its own complex star formation history (SFH)
\citep[e.g.][]{grebel03}. However, in contrast to dIrrs, dSphs are
composed mainly of old and intermediate-age stars, and do not contain
young stellar populations \citep[e.g.][]{holtzman06}. We will test whether
this difference is reflected in the GC populations of our sample
field/group dwarfs and investigate the chemical compositions of their GC
stellar populations.

\begin{figure*}
\centering
\includegraphics[width=12cm,bb=18 148 597 776]{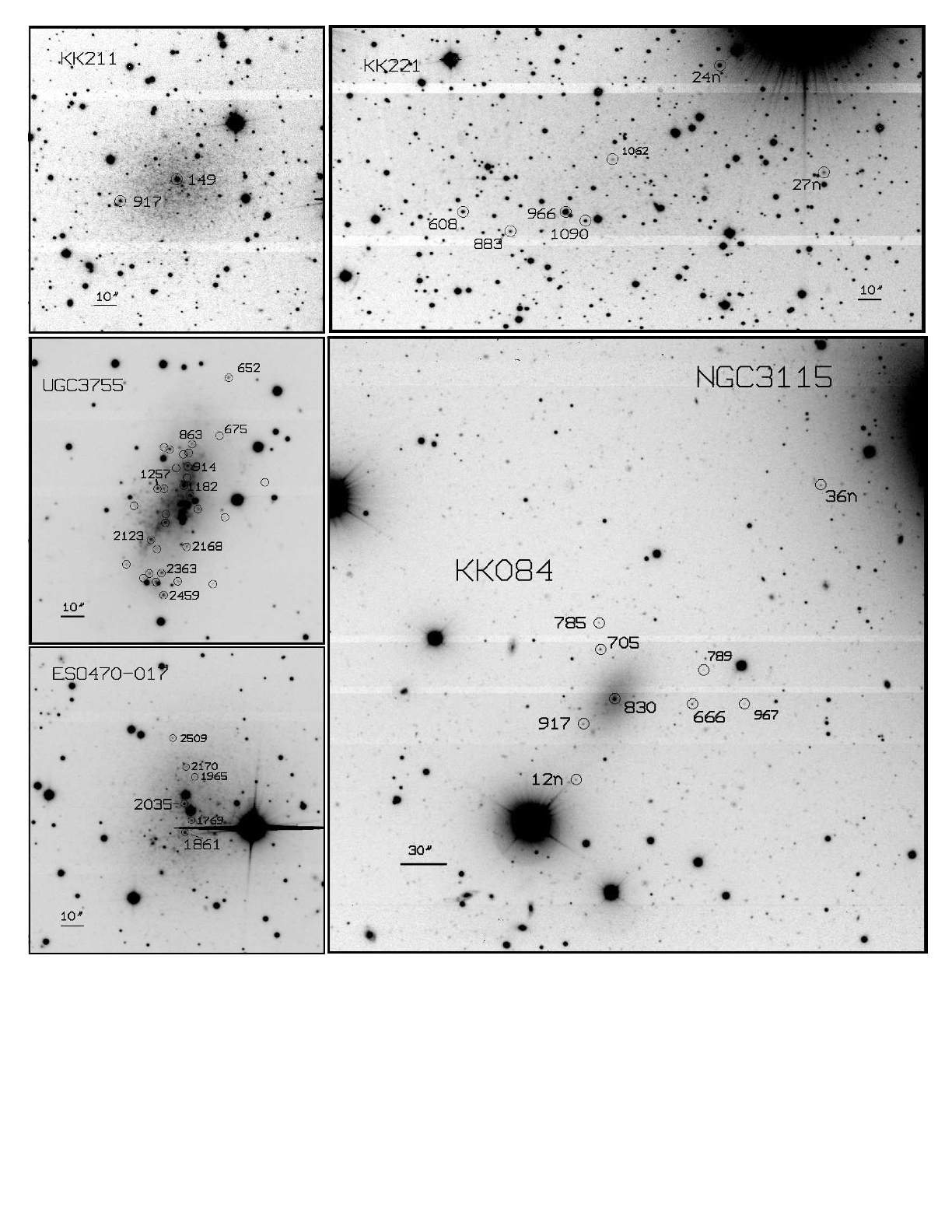}
\caption{FORS-2 images of our sample dwarf galaxies with marked GCs. North is 
to the top, East is on the left. The objects are labeled according to Table~\ref{gcprop}.}
\label{images}
\end{figure*}

\subsection{Pre-imaging data}
All pre-imaging data were obtained with the FORS2 instrument at UT1 (unit
telescope 1, ANTU) as par of the program 76.B-0137 (see also
Fig.~\ref{images}). A journal of the pre-imaging observations is provided
in Table~\ref{log}. All images were reduced using standard techniques
(bias subtraction and flat fielding). We used the FIMS (FORS Instrumental
masks simulator) task {\it fsmosaic} to merge two files of each image. To
register and combine the six sub-integrations obtained in each filter we
used the tasks {\it ccmap} and {\it imcombine} in IRAF\footnote{IRAF is
distributed by the National Optical Astronomy Observatories, which are
operated by the Association of Universities for Research in Astronomy,
Inc., under cooperative agreement with the National Science Foundation.}.
Aperture photometry of GC candidates (GCCs) was performed using the PHOT
task of DAOPHOT-II \citep{stetson87} package implemented in MIDAS. The
detection threshold was set to 3-$\sigma$ above the background. The
minimum full width at half maximum (FWHM) input parameter was
$\sim\!0.76$\arcsec, corresponding to a typical stellar FWHM. To convert
instrumental magnitudes into the Johnson-Cousins standard system we
applied the FORS2 photometric
zeropoints\footnote{http://www.eso.org/observing/dfo/quality/FORS2/qc/
\\zeropoints/zeropoints.html}:
$B_0 = 27.356 \pm 0.007$ (for chip 1), $B_0 = 27.338 \pm 0.005$ (for chip
2), $I_0 = 27.555 \pm 0.1$ (for chip 1), $I_0 = 27.559 \pm 0.07$ (for chip
2), obtained from data taken during the same nights. Atmospheric extinction
coefficients\footnote{http://www.eso.org/observing/dfo/quality/FORS2/qc/
\\photcoeff/photcoeffs\_fors2.html} were taken for 2006-01-01, $k_B=0.269
\pm 0.016$ (for chip 1), $k_B=0.22 \pm 0.22$ (for chip 2), $k_I=0.150 \pm
0.018$ (for chip 1), $k_I=0.135\pm 0.015$ (for chip 2). Finally, the
magnitudes of all GCCs were corrected for Galactic extinction using
reddening maps from \cite{sch98}. The accuracy of our photometry
depends primarily on the accuracy of the background estimates. The
uncertainties grow in the central regions of galaxies where the background
is less homogeneous, and the images become increasingly affected by
stellar crowding. In general, the errors of $B$ and $I$ magnitudes are
less than $0.1$ mag for objects brighter than $21.5$ mag. The budget of
errors includes the errors of aperture photometry, and the uncertainties
of transformations into the standard $B$ and $I$ system. We consider the
internal reddening within our sample galaxies to be $E_{(B-V)}\la0.1$ mag
\citep{james05}, given the similarity of these systems to nearby low
surface brightness galaxies.

For our spectroscopic observations we selected GCCs with integrated colors
$0.7\!<\!(B-I)_0\!<\!2.3$, similar to the selection of Puzia et al.~(2004,
hereafter: P04). In total we found 96, 74, 14, 19, and 11, objects in and
around KK084, UGC3755, ESO490-017, KK221, and KK211, respectively. The
full list of GCCs detected on the VLT images with equatorial coordinates,
$B$ and $I$ magnitudes, and spectroscopic classification of genuine GCs,
foreground stars, and background galaxies is available upon request from
the authors (see also Tabs.~\ref{budget} and \ref{gcprop}). We point out that all
GCCs that are listed in SPM05 are also included in the final target list
and were all selected purely based on their $B\!-\!I$ colours.

To obtain accurate estimates of specific frequencies, we also perform
surface photometry of our sample galaxies using the {\sc SURPHOT} routine
implemented in MIDAS. All steps are identical to those described in
\cite{mak99}. Table~\ref{dwgprop} documents the results, which are
illustrated in Figure~\ref{surf} and discussed in detail in
Appendix~\ref{sbprof}.

\begin{deluxetable}{lcclcc}
\tabletypesize{\scriptsize}
\tablecaption{Journal of pre-imaging observations\label{log}}
\tablewidth{0pt}
\tablehead{
\colhead{Object} & \colhead{Date} & \colhead{Filter} & 
\colhead{$t_{\rm exp}$} & \colhead{Seeing} & \colhead{Airmass} 
}
\startdata
KK211    & 25.08.2005 & B & 6x180   & 1.0\arcsec & 1.740 \\
	 & 25.08.2005 & I & 6x90    & 1.0\arcsec & 1.834 \\
KK221    & 25.08.2005 & B & 6x180   & 1.1\arcsec & 1.914 \\
	 & 25.08.2005 & I & 6x90    & 1.2\arcsec & 1.997 \\
UGC3755  & 01.11.2005 & B & 6x180   & 0.9\arcsec & 1.258 \\
	 & 01.11.2005 & I & 6x90    & 0.9\arcsec & 1.242 \\
E490-017 & 28.11.2005 & B & 6x180   & 0.7\arcsec & 1.041 \\
	 & 28.11.2005 & I & 6x90    & 0.7\arcsec & 1.056 \\
KK084    & 09.12.2005 & B & 6x215   & 1.2\arcsec & 1.164\\\smallskip
	 & 09.12.2005 & I & 6x120   & 1.2\arcsec & 1.164 
	 \enddata
\tablecomments{The exposure time is given in seconds.}
\end{deluxetable}

\subsection{Spectroscopic data}

\begin{figure}
\centering
\includegraphics[width=8.5cm, bb=15 0 416 792]{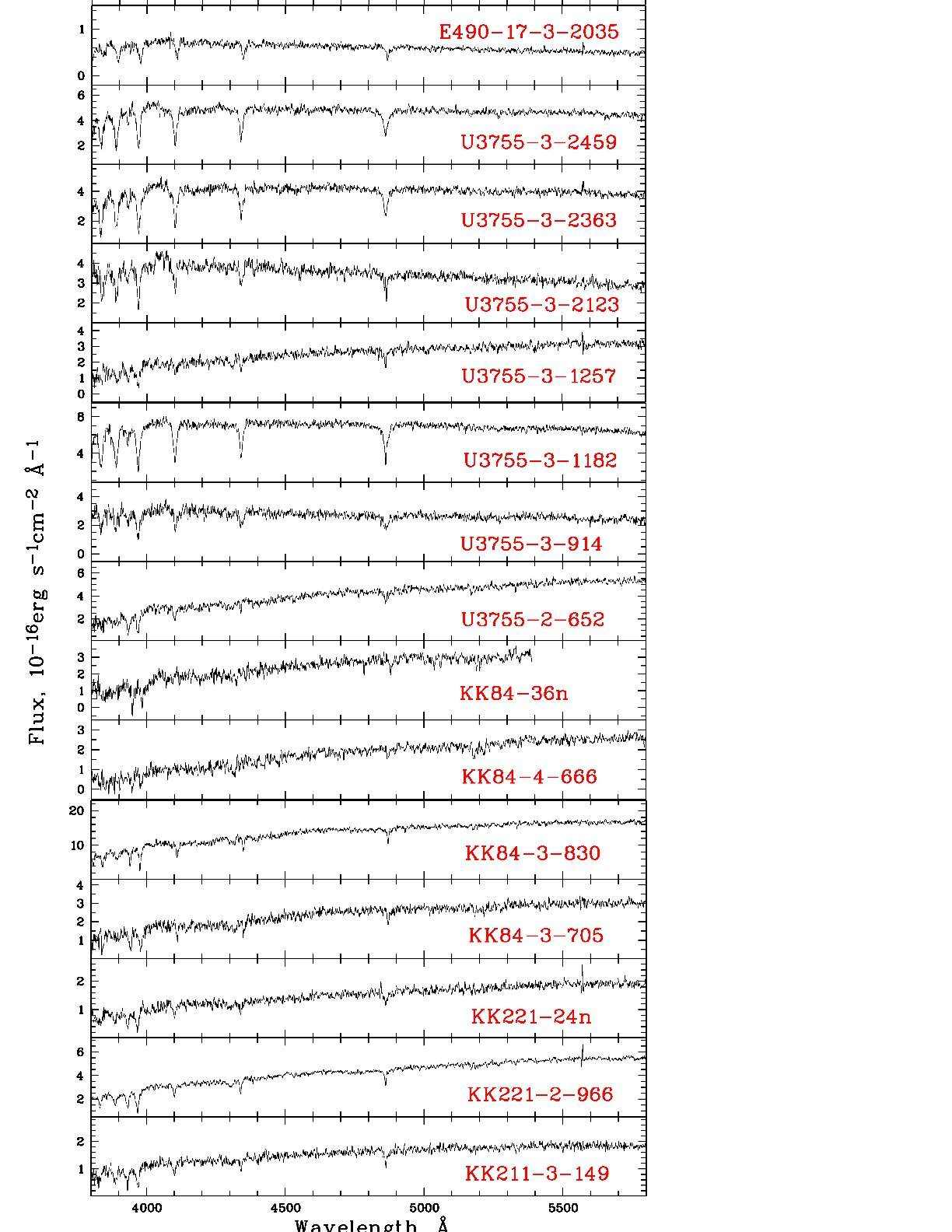}
\caption{Flux calibrated spectra of GCs in our sample dwarf galaxies.}
\label{spec_all}
\end{figure}

\begin{deluxetable}{lccc}
\tabletypesize{\scriptsize}
\tablecaption{Journal of spectroscopic observations \label{splog}}
\tablewidth{0pt}
\tablehead{
\colhead{Object} & \colhead{Date} & 
\colhead{$t_{\rm exp}$} & \colhead{Seeing} 
}
\startdata
E490-017 & 03.01.2006 & 4x1800  & 0.8\arcsec \\
	 & 04.01.2006 &   1800  & 1.6\arcsec \\
KK084    & 03.01.2006 & 5x1800  & 1.0\arcsec \\
	 & 04.01.2006 & 8x1800  & 1.5\arcsec \\
	 & 05.01.2006 & 4x1800  & 0.7\arcsec \\
KK221    & 03.01.2006 & 5x1800  & 1.0\arcsec \\
UGC3755  & 04.01.2006 & 6x1800  & 1.3\arcsec \\
	 & 05.01.2006 & 5x1800  & 0.8\arcsec \\
KK211    & 05.01.2006 & 5x1800  & 0.8\arcsec \\
	 &            &         &     \\
HD013043 & 03.01.2006 & 4, 1    & 1.8\arcsec \\
HR0695   & 03.01.2006 & 2, 1    & 1.8\arcsec \\
	 & 04.01.2006 & 1, 1    & 1.8\arcsec \\
HR1506   & 03.01.2006 & 1, 8    & 1.8\arcsec \\
	 & 05.01.2006 & 1       & 0.8\arcsec \\
HD64606  & 04.01.2006 & 1, 3    & 1.8\arcsec \\
HR3905   & 04.01.2006 & 1, 5    & 1.8\arcsec \\
HR2233   & 05.01.2006 & 1, 1    & 1.3\arcsec \\
HD36003  & 05.01.2006 & 1, 2    & 1.3\arcsec \\\smallskip
HR1015   & 05.01.2006 & 1, 3    & 1.3\arcsec \enddata
\tablecomments{The exposure times for GCs ({\it upper part}) are given in
multiples of seconds, while the exposure times for Lick standard stars 
are shown in seconds for each individual integration.}
\end{deluxetable}

\begin{deluxetable*}{lccccccccr}
\tabletypesize{\scriptsize}
\tablecaption{Resum\'{e} of GC detections. \label{budget}}
\tablewidth{0pt}
\tablehead{
\colhead{Name} & \colhead{GCCs} & \colhead{Obj$_{\rm sel.}$} & \colhead{\#Slits} & 
\colhead{\#Slits$_{\rm SPM05}$} & \colhead{GCs} & \colhead{Gal.} & \colhead{Stars} & 
\colhead{Faint} & \colhead{faint GCCs}
}
\startdata
KK211   		& 2         & 11         &26      &   2      &  2         &     5     &  12  &  7     &      \\
KK221   		& 5         & 19         &36      &   5      &  6         &     6     &   5   & 19    &KK221-3-1062      \\
KK084   		& 8         & 96         &39      &   7      &  7         &    13    &   7   & 12    &KK84-2-789       \\
UGC3755 	& 32       & 74         &39      &  10     & 10        &     8     &   9   & 12    &U3755-3-1963     \\
E490-017	& 5         & 14         &28      &   3      &  2         &     9     &   8   & 9      &E490-017-3-1861  \smallskip
\enddata
\tablecomments{Columns contain numbers of: (2) GCCs in each galaxy listed 
by SPM05, (3) total number of GCCs selected on the VLT images, (4) slits on GCCs in
total, (5) number of slits on SPM05 targets, (6) spectroscopically
confirmed GCs, (7) number of background galaxies in slits, (8) Galactic
stars, (9) faint object in total, the nature of which is unclear. In the
last column we list the names of GCs that are too faint to measure their
radial velocities with reliable accuracy from our observations.}
\end{deluxetable*}

The spectroscopic data were obtained in the MXU mode with FORS2 using
custom slit masks for KK084, UGC~3755, ESO~490-017, KK221, and KK211 that
contain 39, 39, 28, 36, and 26 objects in total, respectively. A journal
of spectroscopic observations is provided in Table~\ref{splog}. The masks
included our primary target GCCs, as well as mask-filler objects that were
mostly stars and background galaxies. In general, due to the concentrated
location of GCCs in the central regions of dwarf galaxies we primarily
targeted GCCs from SPM05 and set slits on:
all two GCCs in KK211, all five GCCs in KK221, 3 of 5 GCCs in ESO490-017,
10 of 32 GCCs in UGC3755, and 7 of 8 GCCs in KK084 (KK084-2-974 was not
observed, see also Tables~\ref{budget} and \ref{gcprop} for details).

The reduction of the spectroscopic data and the subsequent analysis were
performed with a combination of MIDAS and IRAF tasks. After cosmic-ray
removal and bias subtraction all frames were divided by a normalized
flat-field image. For each slit a 2-dimensional subsection of the CCD was
extracted and then treated separately. To correct the effect of optical
field distortions in the FORS field-of-view we applied the method
described by P04, which is based on a two-dimensional wavelength solution
assembled of 1-D solutions for each pixel row from arclamp spectra. For
this task we used the {\sc LONG} package in MIDAS. This procedure requires
an accurate wavelength solution for each individual pixel row. To maximize
the signal-to-noise ratio (S/N) of lines in the comparison arc spectrum we
flat-fielded the arc spectrum to correct for small-scale fluctuations and
applied a median filter to the frame using a rectangular filter window of
one pixel in the dispersion direction and 3 pixels in the spatial
direction. A typical accuracy of the 2-dimensional dispersion solution was
$\le\!0.2$ \AA.~The extraction region was defined by tracing the GC
spectrum along the wavelength dimension. The window width for the traced
extraction was set so that the boundaries of spectra were at $\sim\!15$\%
of the peak flux in all cases, except for GCs 1182 and 2123 in UGC~3755
where the extraction window was set to a width of $\sim\!20$\% of the peak
flux, because of their location near regions of $H\alpha$ emission.

To ensure that the dispersion solution was determined correctly we
extracted the 1-dimensional spectra using the method described above
without subtracting the sky spectrum and determined the wavelengths of the
telluric lines \citep{ost1996, ost2000}.~For few spectra we found
systematic shifts of the order of $\la\!1$\AA\ equal for all lines. Such
shifts were corrected by adding of the corresponding correction term to
the dispersion solution. The GC spectra are then cross-correlated with
spectra of radial-velocity standard stars observed in the same night using
the {\sl xcorrelate/image} procedure in MIDAS, which yields radial
velocities according to the method of \cite{tonry79}. Table~\ref{gcprop}
summarizes the measured heliocentric radial velocities.

\begin{figure}
\centering
\includegraphics[width=8cm]{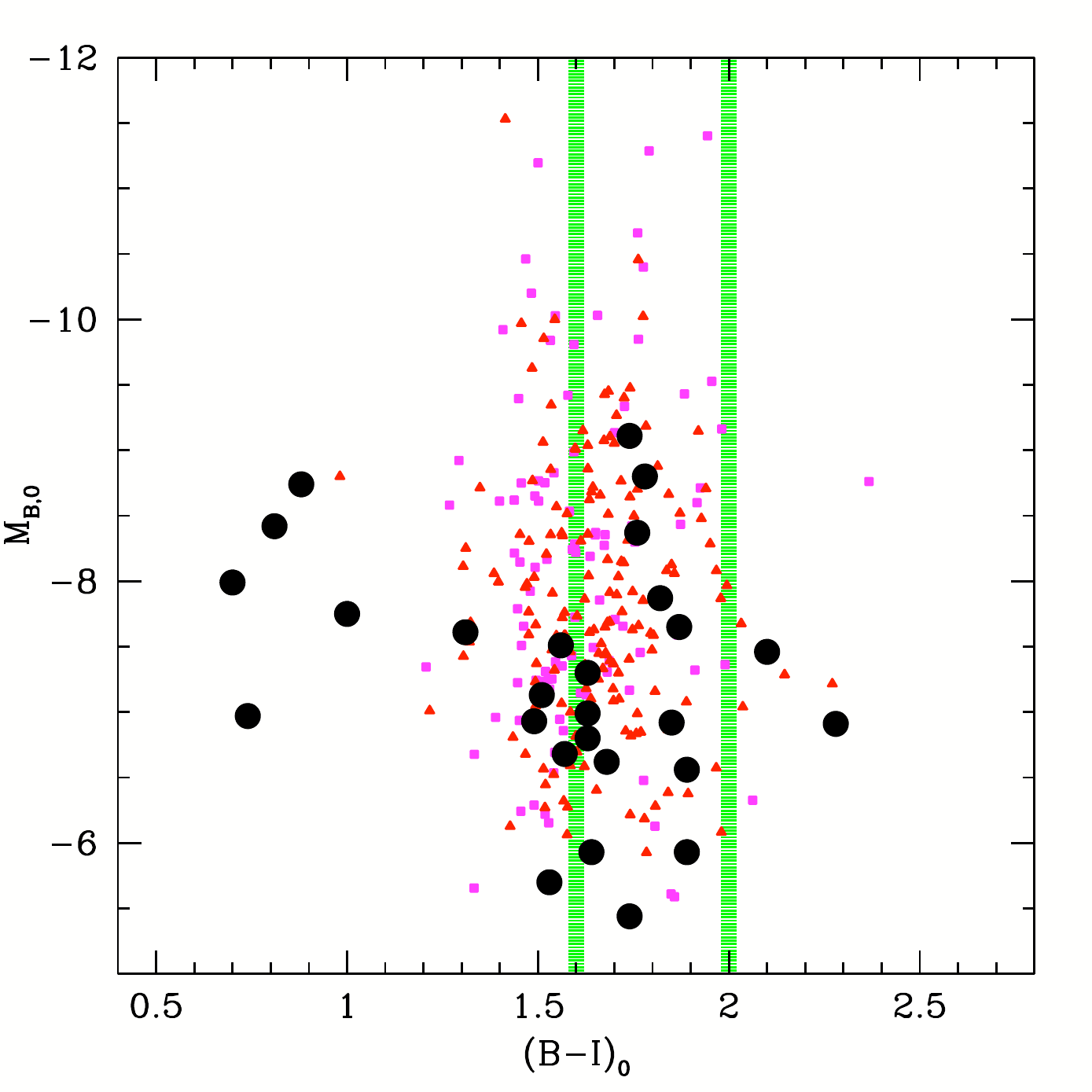}
\caption{Color-magnitude diagram of confirmed GCs in LSB dwarf galaxies 
({\it large circles}). Overplotted are also Milky Way ({\it small squares}) and M31
GCs ({\it small trinalges}). The photometry for Milky Way GC was taken
from the McMaster catalog \citep{harris91}, while the M31 data was adopted
from \cite{barmby00}. The two vertical lines indicate the location of the
blue and red GC sub-populations in giant elliptical galaxies
\citep[e.g.][]{p04}.}
\label{gcphotcmd}
\end{figure}

\begin{deluxetable*}{lcccccr}
\tabletypesize{\scriptsize}
\tablecaption{Properties of spectroscopically confirmed GCs
in our sample LSB dwarf galaxies\label{gcprop}}
\tablewidth{0pt}
\tablehead{
\colhead{GC} & \colhead{RA (J2000) DEC} & \colhead{$B$} & \colhead{$M_B$} & 
\colhead{$B-V$} & \colhead{$B-I$} & \colhead{$V_h$} 
}
\startdata
KK211-3-149     &  13 42 05.6 $-$45 12 18   & 20.64   & $-7.13$ & 0.69   & 1.51 & 580$\pm$23  \\
KK211-3-917     &  13 42 08.0 $-$45 12 28   & 21.84   & $-5.93$ & 0.93   & 1.64 & 620$\pm$39  \\
KK221-2-608     &  13 48 55.1 $-$47 00 07   & 21.08   & $-6.92$ & 1.12   & 1.85 & 541$\pm$32 \\
KK221-2-883     &  13 48 53.0 $-$47 00 16   & 22.07   & $-5.93$ & 1.14   & 1.89 & 546$\pm$46 \\
KK221-2-966     &  13 48 50.5 $-$47 00 07   & 19.20   & $-8.80$ & 1.00   & 1.78 & 509$\pm$25 \\
KK221-2-1090   &  13 48 49.6 $-$47 00 11   & 21.20   & $-6.80$ & 0.97   & 1.63 & 478$\pm$29 \\
KK221-24n        &  13 48 43.6 $-$46 58 59   & 20.70   & $-7.30$ & \nodata& 1.63 & 512$\pm$31 \\
KK221-27n        &  13 48 39.0 $-$46 59 49   & 22.26   & $-5.44$ & \nodata& 1.74 & 466$\pm$35 \\
KK084-2-785     &  10 05 35.8 $-$07 44 06   & 23.31   & $-6.62$ & 0.68   & 1.68 & 856$\pm$24 \\
KK084-3-705     &  10 05 35.7 $-$07 44 25   & 22.28   & $-7.65$ & 0.73   & 1.87 & 670$\pm$31 \\
KK084-3-830     &  10 05 35.0 $-$07 44 59   & 20.82   & $-9.11$ & 0.57   & 1.74 & 594$\pm$32 \\
KK084-3-917     &  10 05 36.5 $-$07 45 16   & 22.94   & $-6.99$ & 0.53   & 1.63 & 619$\pm$28 \\
KK084-4-666     &  10 05 31.5 $-$07 45 03   & 22.47   & $-7.46$ & 0.91   & 2.10 & 678$\pm$21 \\
KK084-12n        &  10 05 36.8 $-$07 45 54   & 23.00   & $-6.93$ & \nodata& 1.49 & 911$\pm$40 \\
KK084-36n        &  10 05 25.6 $-$07 42 33   & 23.02   & $-6.91$ & \nodata& 2.28 & 1210$\pm$27\\
UGC3755-2-652     &   07 13 50.1 +10 32 15   & 21.48   & $-7.87$ & 1.11   & 1.82 & 323$\pm$21 \\
UGC3755-2-675     &   07 13 50.4 +10 31 49   & 23.65   & $-5.70$ & 0.81   & 1.53 & 367$\pm$21 \\
UGC3755-2-863     &   07 13 51.3 +10 31 45   & 22.79   & $-6.56$ & 1.13   & 1.89 & 290$\pm$33 \\
UGC3755-3-914     &   07 13 51.4 +10 31 35   & 21.74   & $-7.61$ & 0.75   & 1.31 & 284$\pm$24 \\
UGC3755-3-1182    &  07 13 51.5 +10 31 26   & 20.61   & $-8.74$ & 0.56   & 0.88 & 335$\pm$32 \\
UGC3755-3-1257    &  07 13 52.3 +10 31 24   & 20.98   & $-8.37$ & 1.10   & 1.76 & 327$\pm$31 \\
UGC3755-3-2123    &  07 13 52.5 +10 31 01   & 21.36   & $-7.99$ & 0.53   & 0.70 & 329$\pm$22 \\
UGC3755-3-2363    &  07 13 52.2 +10 30 45   & 21.60   & $-7.75$ & 0.76   & 1.00 & 312$\pm$18 \\
UGC3755-3-2168    &  07 13 51.4 +10 30 58   & 21.84   & $-7.51$ & 0.88   & 1.56 & 324$\pm$28 \\
UGC3755-3-2459    &  07 13 52.2 +10 30 35   & 20.93   & $-8.42$ & 0.32   & 0.81 & 333$\pm$32 \\
E490-017-3-2035  & 06 37 57.3 $-$25 59 59   & 21.16   & $-6.97$ & 0.33   & 0.74 & 529$\pm$34 \\
E490-017-3-1861  & 06 37 57.3 $-$26 00 13   & 21.45   & $-6.68$ & 0.50   & 1.57 & 522$\pm$9    \smallskip
\enddata
\tablecomments{Columns contain the following data: (2), (3) equatorial
coordinates, (4) integrated $B$ magnitude from our FORS2 photometry
corrected for Galactic extinction \citep{sch98}, (5) absolute magnitude computed 
with the distances from Table~\ref{dwgprop}, (6),(7) integrated
$B\!-\!V$ and $B\!-\!I$ colours corrected for Galactic extinction
\citep{sch98}, (8) heliocentric radial velocities measured in this study.}
\end{deluxetable*}

\subsection{Detection Efficiencies}

\begin{figure*}
\centering
\includegraphics[width=14cm,bb=47 204 612 657]{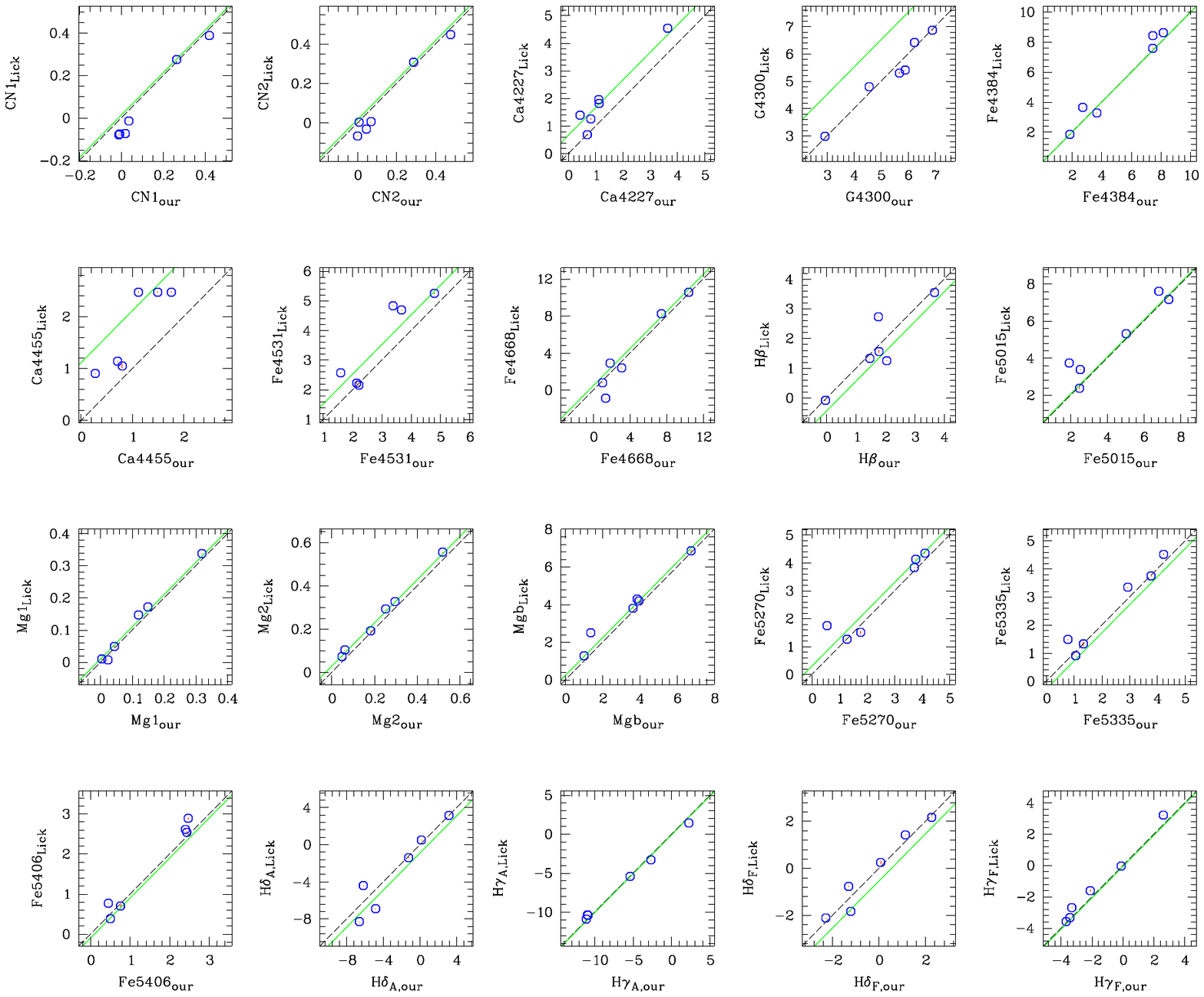}
\caption{Comparison of passband measurements of our spectra and original Lick
data for 10 Lick standard stars. The dashed line shows the one-to-one relation.
The grey line shows the relation from P04.}
\label{compar}
\end{figure*}

We present a resum\'{e} of GC detection efficiencies for our spectroscopic
observations in Table~\ref{budget}. Numbers of detected GCCs within the
FORS2 field-of-view with colors resembling those of GCCs, prepared masks
for a part of them, confirmed genuine globular clusters, distant galaxies
with radial velocities $\ge\!1000$ km s$^{-1}$ , and Galactic stars with
radial velocities $\sim\!0$ km s$^{-1}$ are given for each galaxy,
correspondingly. All observed GCCs from SPM05 appear to be genuine
globular clusters, except a distant galaxy KK084--4--967. Heliocentric
radial velocities of spectroscopically confirmed GCs are similar to the
system velocities of Cen~A and NGC3115 in the cases of KK211, KK221, and
KK84, and to the velocities of the host galaxies measured using indendent
methods in the cases of UGC3755 and E490-017 (Table~\ref{gcprop}). The
radial velocity dispersion of the UGC~3755 sample is small $\sim\!10$ km
s$^{-1}$, and the mean is consistent with the systemic value $V_h=315$ km
s$^{-1}$ measured by \cite{bicay86}. UGC~3755 is the only galaxy in our
sample with a low-enough systemic radial velocity so that a test of
contamination likelihood by Galactic foreground stars is indicated. An
evaluation of this likelihood with the stellar population synthesis models
of the Milky Way \citep{robin03} shows that the expected radial velocities
of the thick disk, spheroid, and bulge component in the direction of
UGC~3755 are significantly below $200$ km/s. Combined with their diffuse
(i.e. non-stellar) PSFs in the HST images and their color information all
our GCCs in UGC~3755 with accurate radial velocity measurements (see
Tab.~\ref{gcprop}) are highly unlikely to be foreground stars. We
discovered three new GCs: KK84-36n and KK84-12n, and KK221-12n (see
Fig.1). The equatorial coordinates, total $B$ magnitudes, absolute $B$
magnitudes, $B-V$, $B-I$ colors and radial velocities for all GCs
confirmed or detected in this work are given in Table~\ref{gcprop}.
In summary, our GCC selection efficiency based on the HST imaging survey
presented in SPM05 is higher than 96\% for the observed sample.

The distribution of confirmed GCs in the color-magnitude diagram is
illustrated in Figure~\ref{gcphotcmd} where we compare our sample with the
distribution of Milky Way and M31 GCs.~The bulk of our sample has
$B\!-\!I$ colors very similar to the Local Group GCs.~A few GC have
relatively blue colors which suggests younger ages. Our sample GCs probe
the luminosity range around the turn-over magnitude of the GC luminosity
function (GCLF: The turn-over for old GC populations is expected at
$M_B\approx-7.1$, assuming $M_{V}\simeq-7.66\pm0.11$ from Di Criscienzo et
al. [2006] for metal-poor Milky Way GCs and a typical
$B\!-\!V\approx0.5-0.7$ for metal-poor stellar populations from Bruzual \&
Charlot [2003] SSP models) down to about a factor $2-3$ fainter clusters.
We recall that SPM05 found indications for an excess cluster population in
this magnitude range \citep[see also][]{vdb07, jordan07} and we point out
that a significant number of GC candidates from the SPM05 sample are being
confirmed as genuine GCs at these faint magnitudes. However, a larger
sample is required to robustly quantify the excess of clusters with
respect to GC systems of more massive galaxies, such as the two Local
Group spirals.

\section{Analysis}
\label{analysis}
\subsection{Lick Index Measurement and Calibration}
\label{calibr_Lick}

\begin{figure*}
\centering
\includegraphics[width=7.5cm]{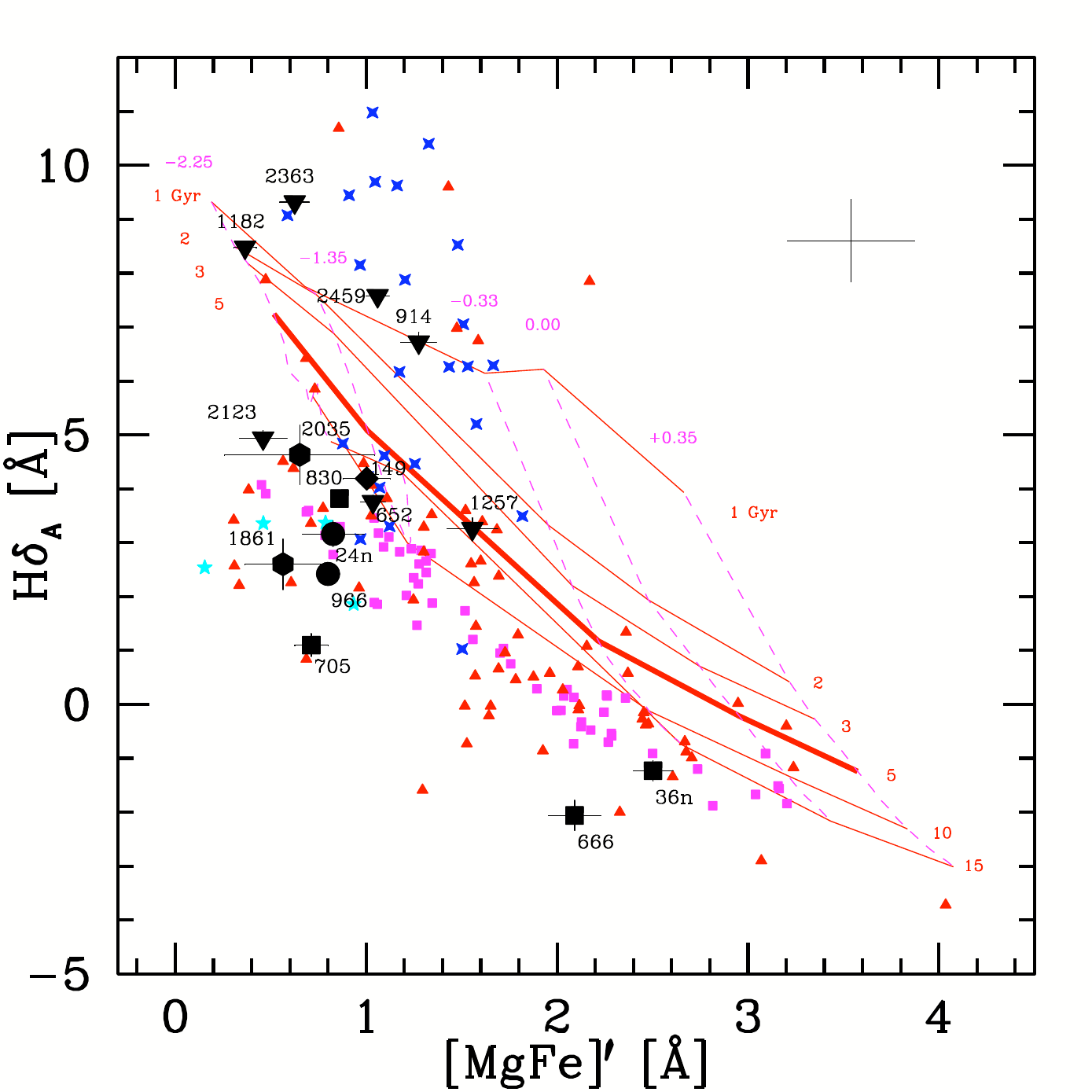}
\includegraphics[width=7.5cm]{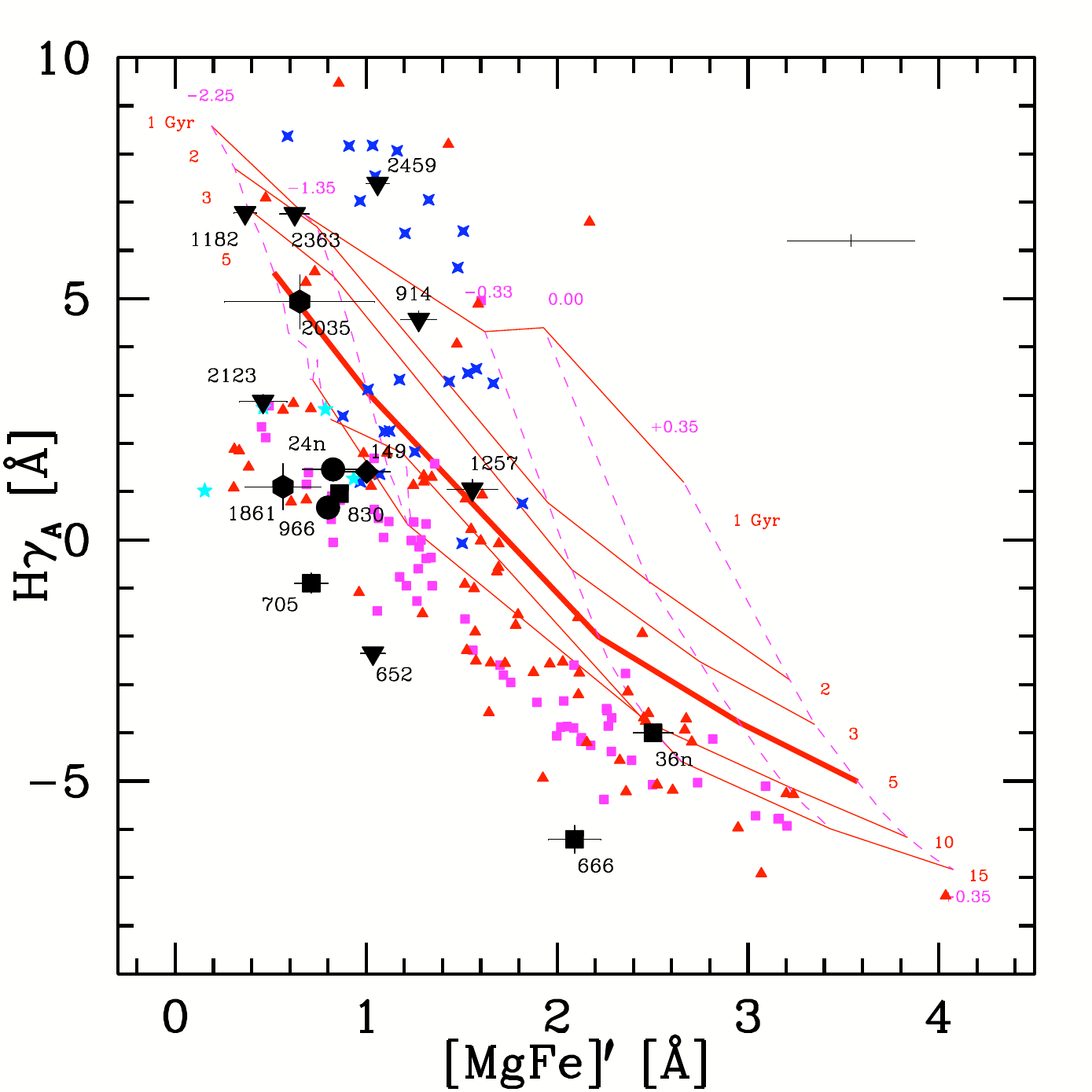}
\includegraphics[width=7.5cm]{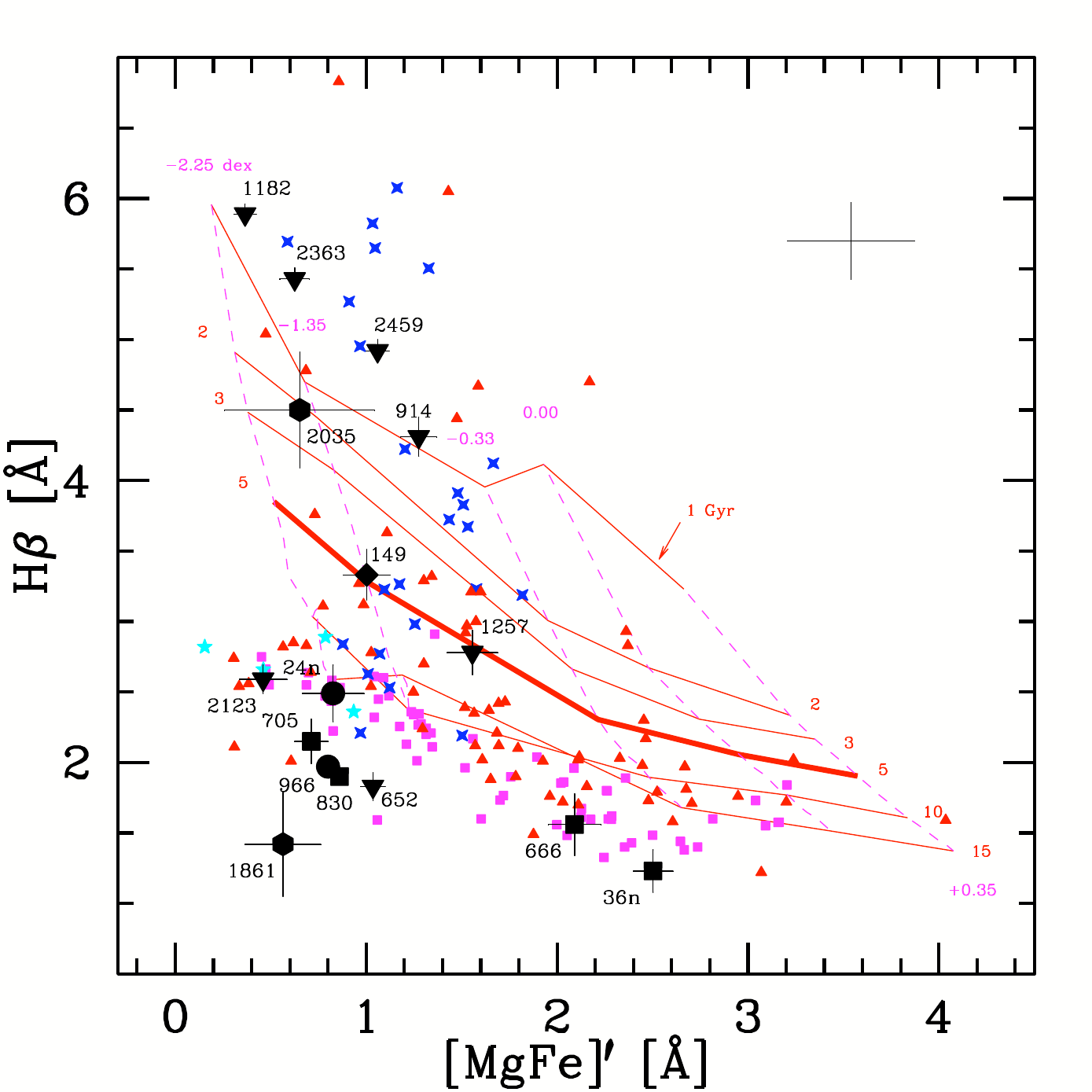}
\includegraphics[width=7.5cm]{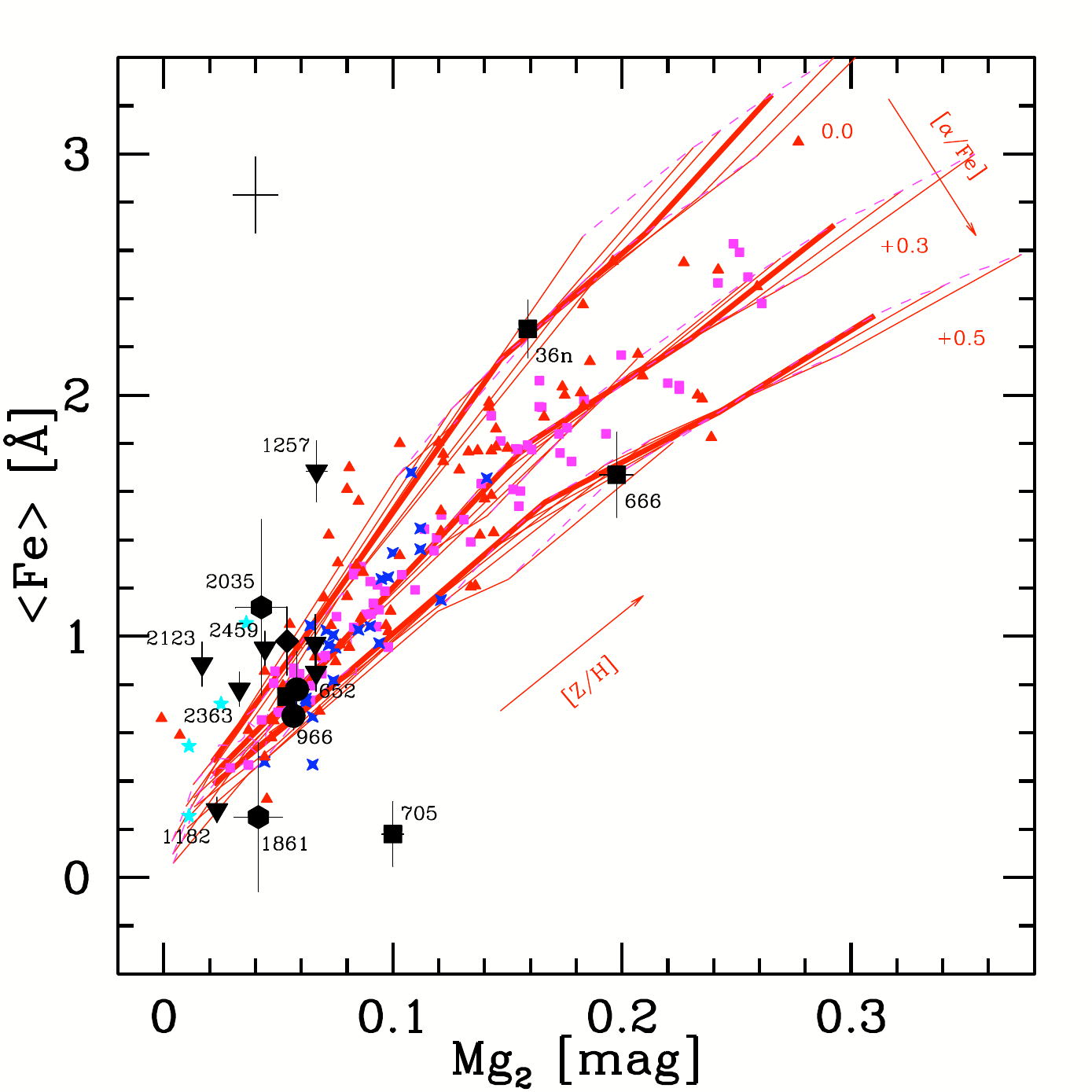}
\caption{Diagnostic plots for globular clusters in KK084 (squares), GC 
149 KK211 (diamond), GCs 966 and 24n in KK221 (circles), GCs 2035 and 1861
in E490-17 (hexagons), and GCs in UGC~3755 (inverted tirangles). We use
SSP model predictions from Thomas et al.~(2003, 2004). The Balmer line
index diagnostic plots (both upper panels and the lower left panel) show
the grids for \afe=0.3 dex. The cross in the corner of each panel
indicates the systematic calibration uncertainty to the Lick index system.
The error bars of individual GCs are the total statistical uncertainties.
Overplotted are Lick index measurements for GCs in the Milky Way
\citep[{\it small squares};][]{p02, schiavon05}, M31 \citep[{\it small
triangles};][]{p05a}, the Large Magellanic Cloud \citep[{\it small 4-prong
stars};][]{beasley02}, and the Fornax dSph galaxy in the Local Group
\citep[{\it small 5-prong stars};][]{strader03a}.}
\label{diagd}
\end{figure*}

\begin{table}
\centering
\tabletypesize{\scriptsize}
\caption{Correction terms of the transformation to the Lick/IDS
standard system \citep{worthey94, worthey97}.}
\label{correction}
\begin{tabular}{lrcc} \\ \hline \hline
Index       & c          & rms error  & Units \\ \hline
CN$_1$      & $-$0.067     & 0.028      & mag    \\
CN$_2$      & $-$0.050     & 0.039      & mag    \\
Ca4227      & 0.487      & 0.247      &  \AA   \\
G4300       & 0.049      & 0.608      &  \AA   \\
Fe4384      & 0.041     & 0.542      &  \AA   \\
Ca4455      & 0.419      & 0.380      &  \AA   \\
Fe4531      & 0.322      & 0.762      &  \AA   \\
Fe4668      & 0.685      & 0.782      &  \AA   \\
H$\beta$    & $-$0.183      & 0.277      &  \AA   \\
Fe5015      & 0.460      & 0.640      &  \AA   \\
Mg$_1$      & 0.010      & 0.007      &  mag    \\
Mg$_2$      & 0.027      & 0.010      &  mag    \\
Mgb         & 0.351      & 0.134      & \AA  \\
Fe5270      & 0.345      & 0.203      & \AA  \\
Fe5335      & $-$0.156     & 0.230      & \AA  \\
Fe5406      & $-$0.208     & 0.124      & \AA  \\
H$_{\delta A}$& $-$0.109    & 0.772      & \AA  \\
H$_{\gamma A}$& $-$0.000    & 0.129      & \AA  \\
H$_{\delta F}$& 0.013    & 0.229      & \AA  \\
H$_{\gamma F}$& 0.491     & 0.148      & \AA  \\
\hline
\end{tabular}
\end{table}

We measure Lick indices with the routine described in \cite{p02} and P04
for GCs with spectra with S/N$\ge\!30$ per \AA.~In general, we select
high-quality spectra for the subsequent analysis of evolutionary
parameters, but include in some interesting cases a few lower-S/N spectra.

P04 calibrated the instrumental FORS/MXU system of Lick line indices into
the standard one using a set of 31 Lick standard stars. The correction
functions were calculated in the form $$ I_{cal} = I_{raw}+ \alpha,$$
where $ I_{cal}$ and $I_{raw}$ are the calibrated and the measured
indices, respectively. To check the correspondence of our index
measurements to the Lick standard system we performed the same analysis
using 10 stars observed together with our target sample. It should be
noted that all standard star spectra were observed with the same slit size
(1\arcsec) and were extracted and smoothed in the same way as the GCs of
our study and as it was done by P04. We show comparison of passband
measurements of our spectra and original Lick data for our sample Lick
standard stars in Figure~\ref{compar}. The dashed line shows the
one-to-one relation. The dotted line shows the relation from P04.
Table~\ref{correction} summarises the coefficients of transformation into
the Lick standard system and the rms of the calibration. Comparison of our
calibration coefficients with the ones form P04 shows that in all cases
but one the transformations agree well within the errors. A large
difference exists for the index G4300. However, P04 noticed that this
index is very noisy and its calibration uncertain. It is worth to note
that the rms errors of transformations into the Lick system are large for
some other indices (e.g. Ca4455, see Table~\ref{correction} this paper and
Table~8 in P04). However, the most important indices for our analysis,
such as H${\beta}$, H$_{\gamma A}$, H$_{\delta A}$, Mg$_1$, Mg$_2$, and
Mg$b$, Fe5270, and Fe5335 have all very robust Lick system calibrations
which agree well with previous work. The calibrated Lick indices for all
confirmed GCs in our sample are summarized in Table~\ref{lickind1}.

\subsection{Ages and Metallicities}
\label{evol_par_GCs}

It was shown in series of papers \citep[see e.g.~P04,][]{p05a, p05b} that
age-metallicity diagnos\-tic plots which make use of different Balmer
indices and the composite metallicity index [MgFe]\arcmin\footnote{This
composite Lick index is defined to be \afe-insensitive,
[MgFe]\arcmin$=\left\{\mbox{Mg}b \cdot (0.72 \cdot \mbox{Fe5270} + 0.28
\cdot \mbox{Fe5335})\right\}^{1/2}$, see \cite{thomas03} for details.}
represent a powerful tool to estimate ages and metallicities of globular
clusters. For data with S/N~$\ga25$ per \AA\ the Balmer index H$\gamma_A$
is most sensitive to age and least sensitive to metallicity and \afe\
variations, as well as the degeneracy between these parameters
\citep{p05b}.~The SSP model predictions of \cite{thomas03, thomas04} for
stellar populations with well-defined \afe\ ratios provide us with the
option to estimate \afe\ ratio from Mg$_2$ versus $ \langle \mbox{Fe}
\rangle = (\mbox{Fe5270} + \mbox{Fe5335})/2$ diagnostic plots.

In Figure~\ref{diagd} we show age-metallicity diagnostic plots for our
sample globular clusters with the highest-S/N spectra. Almost all clusters
have indices consistent with low metallicities, typically [Z/H]~$\le -1$,
and a wide range of ages and \afe\ ratios. We compare the index
measurements of our sample GCs with those of GCs in the Milky Way
\citep{p02, schiavon05}, M31 \citep{p05a}, and the Large Magellanic Cloud
\citep{beasley02}, and find that most objects in our sample have Balmer
line indices, at a given [MgFe]\arcmin, that are comparable with
metal-poor GCs in the Milky Way and M31 in all diagnostic plots of
Figure~\ref{diagd}. Several GCs in UGC~3755, which show stronger Balmer
indices, resemble the sub-population of young star clusters in the LMC
\citep[e.g.][]{kerber07}.

Iterating between the diagnostic plots in Figure~\ref{diagd} converges to
give accurate metallicity and \afe\ estimates, as well as robust relative
ages that allow one to distinguish between old, intermediate-age, and
young stellar populations.~We refer to this approach as the iterative
technique.~Another way to obtain age, metallicity, and \afe\ estimates is
by means of linear interpolation within this three-dimensional space
defined in the models and the subsequent $\chi^2$ minimization of the
difference between observed and predicted indices \citep{sh06}.~This
technique makes use of multiple Lick indices.

\begin{figure*}[!t]
\centering
\includegraphics[width=5.5cm]{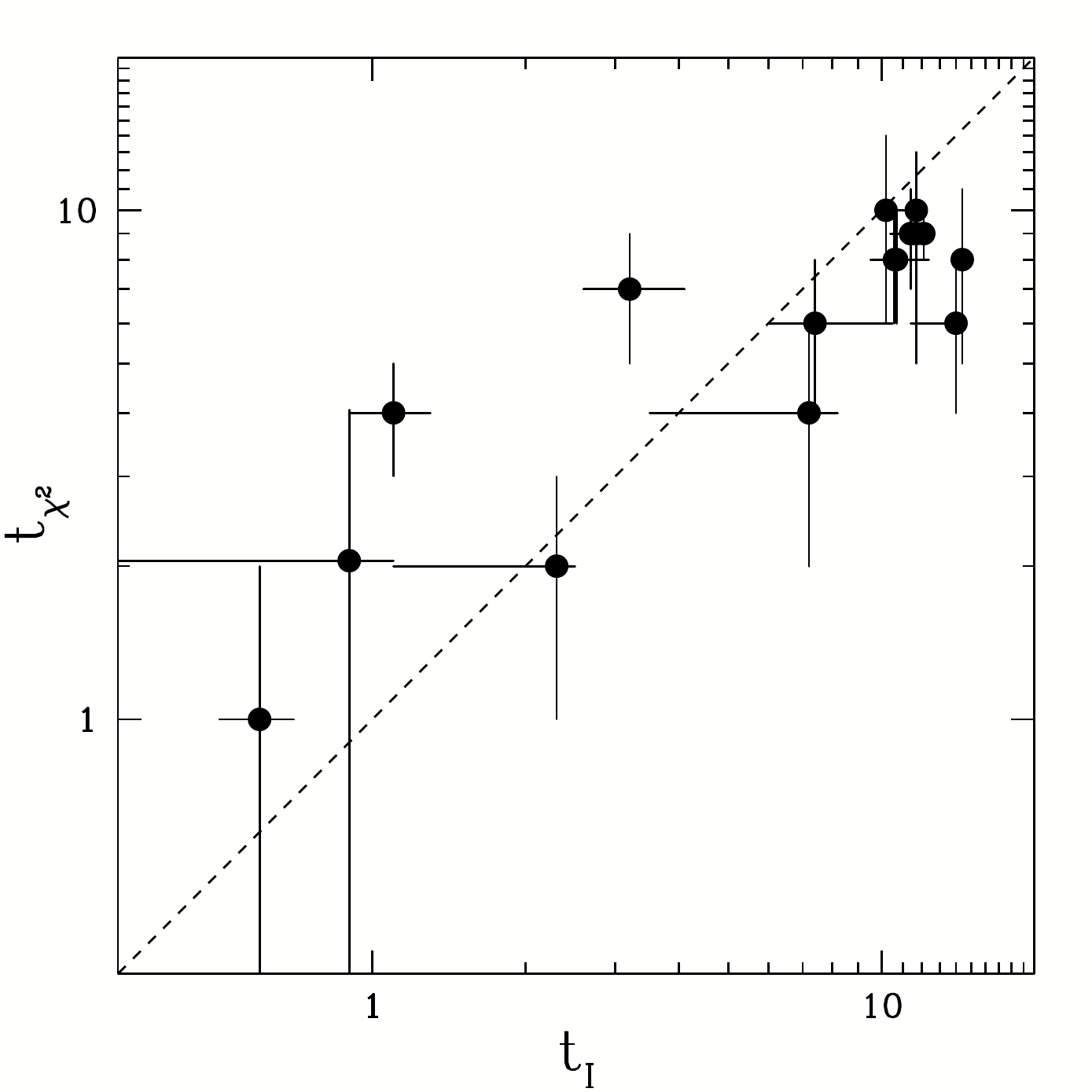}
\includegraphics[width=5.5cm]{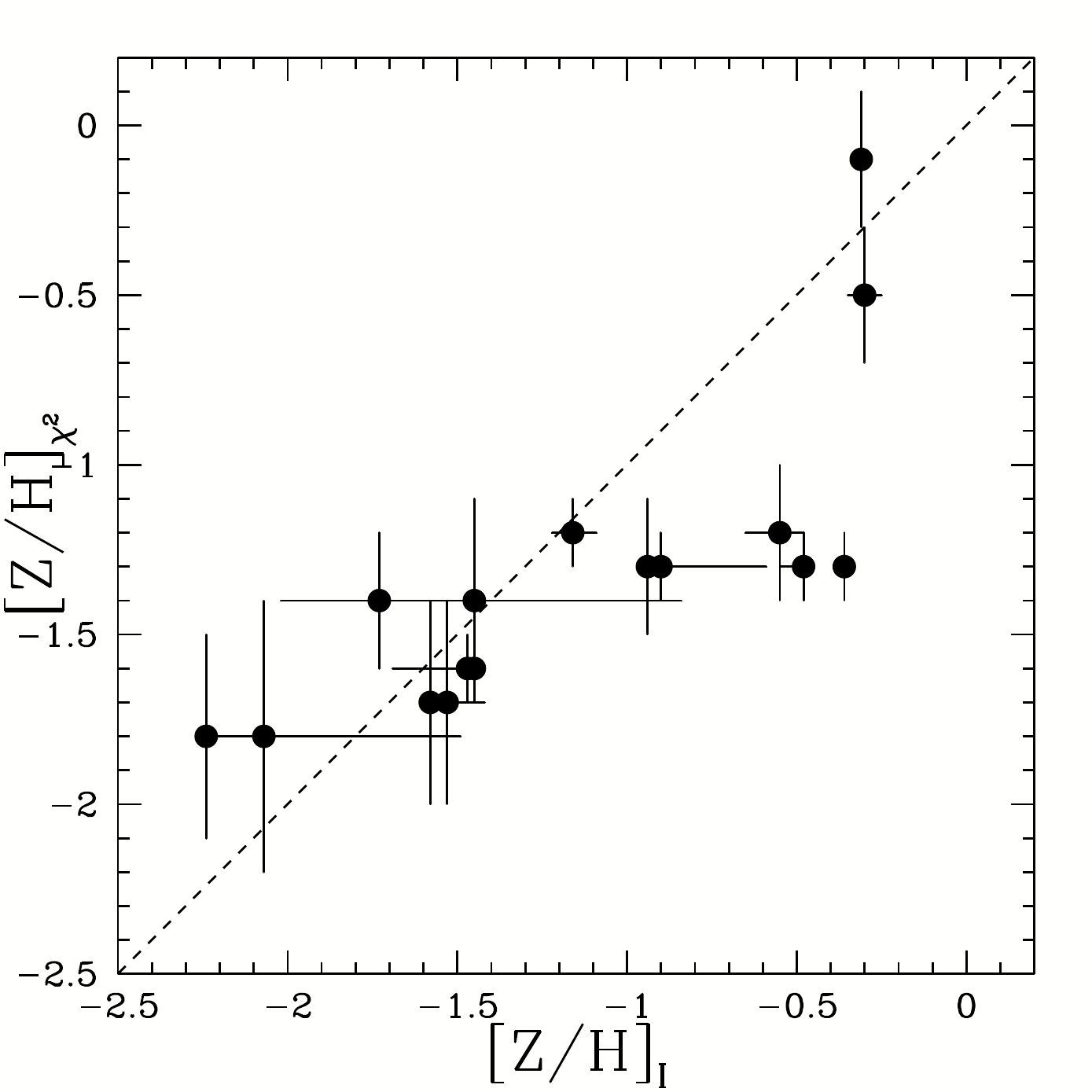}
\includegraphics[width=5.5cm]{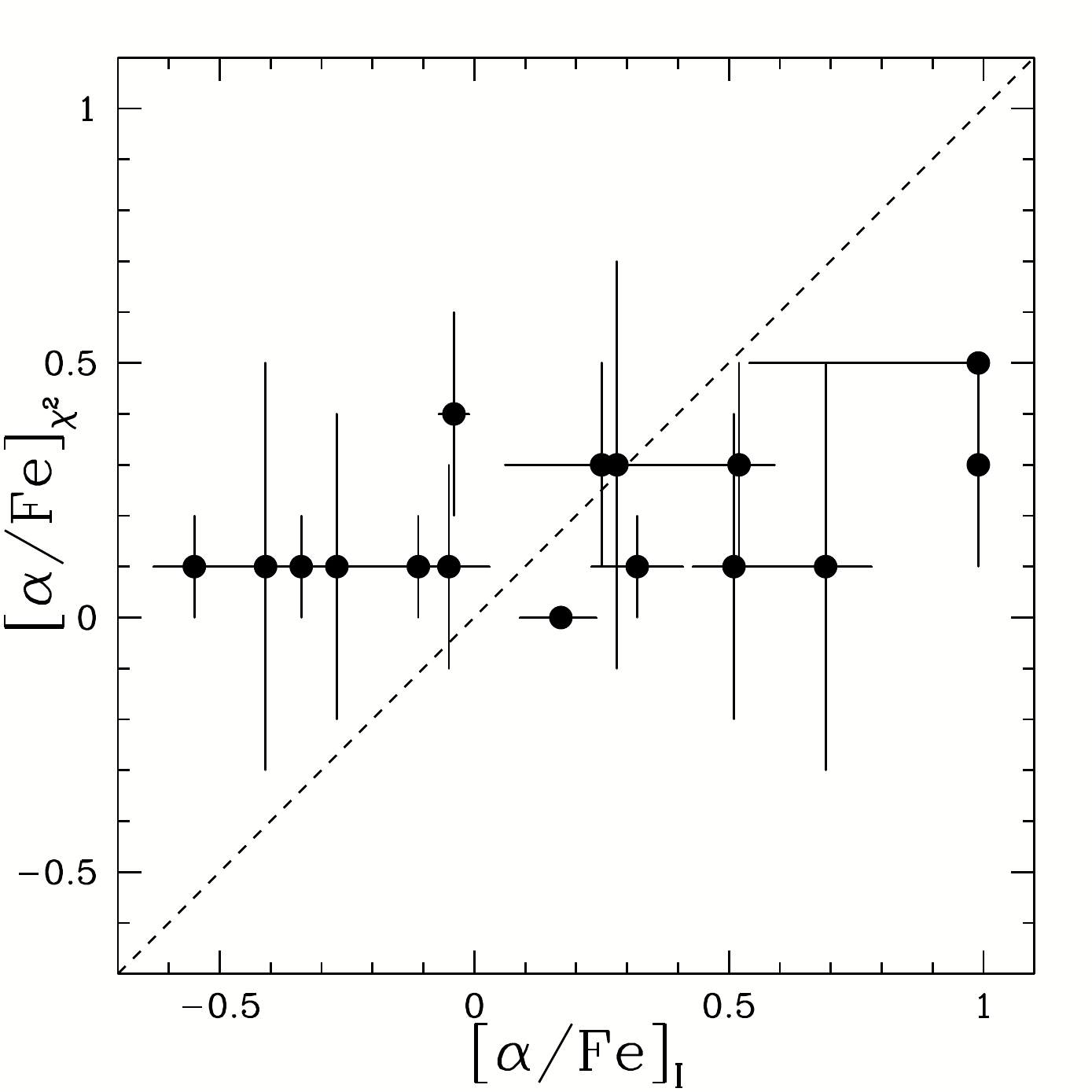}
\caption{Comparison of ages, \zh\, and \afe\ values for our sample GCs 
that were derived using the iterative and $\chi^2$ minimization technique. 
All measurements are summarized in Table~\ref{ssp}.}
\label{amacomp}
\end{figure*}

We apply both techniques to our dataset and summarize the results in
Table~\ref{ssp}. We show a comparison of the output results of the two
techniques in Figure~\ref{amacomp} and find good agreement for age and
metallicity estimates with a scatter about the one-to-one relation that is
consistent with the measurement uncertainties. The $\chi^2$ minimization
technique appears to deliver a limited range of \afe\ values compared to
the iterative approach, which provides results more consistent with the
general distribution of data in diagnostic plots of Figure~\ref{diagd}.
The $\chi^2$ technique works entirely within the parameter space defined
by the model grids. The fact that it does not make use of extrapolations
in the \afe\ grid is likely the reason for the reduced \afe\ dynamic
range.

\subsubsection{GCs in KK084}
\label{agekk84}
All observed globular clusters in this galaxy are old with ages $>8$ Gyr.
The central GC in KK84, KK084-3-830, has the lowest metallicity and the
oldest age among all GCs in this galaxy. We find two metal-poor
[Z/H]~$\approx-1.4$) and two metal-rich clusters [Z/H]~$\approx-0.3$).
Both metal-rich clusters (KK84-36n and KK84-666) are located near NGC~3115
(see Figure~\ref{images}) and have elevated radial velocities, which
grants the possibility that they might be associated with the disk
component of NGC~3115 rather than KK084. Our results for these two GCs are
also in line with the VLT study of GCs in NGC~3115 by \cite{kunt02} and
GMOS spectroscopy data for the diffuse light of NGC~3115 by
\cite{norris06}. \citeauthor{kunt02} found two GC sub-populations with
mean metallicities $\feh \simeq -0.37$ and $-1.36$ dex. The absolute age
was found $11-12$ Gyr for all observed GCs. The long-slit spectroscopy of
the diffuse light in NGC~3115 by \citeauthor{norris06} shows a
luminosity-weighted age of $\sim$6 Gyr for the disk component and a
luminosity-weighted age of $\sim$12 Gyr for the spheroid.

\subsubsection{GCs in UGC~3755}
GCs 1182, 2363, 2459, and 914 in the isolated dwarf irregular galaxy
UGC~3755 appear to be the youngest objects in our sample with ages in
the range $1-4$ Gyr. The other clusters in UGC~3755 are significantly
older. One noteworthy case is U3755-3-1257, which shows an intermediate
age and metallicity. The youngest GC in our sample has the smallest
projected distance relative to the photometric center of the host galaxy.

\subsubsection{GCs in ESO490-17}
The two GCs in our sample that are associated with ESO490-17 are both
metal-poor. Although our data for this galaxy has the lowest S/N, the age
difference between the two GCs with high-quality spectra is seen clearly
in the diagnostic plots of Figure~\ref{diagd}. This difference is
reflected in the integrated colors (see Table~\ref{gcprop}). In terms of
projected distance, the young cluster GC 2035 is located closer to the
galaxy center.

\subsubsection{GCs in KK211}
We find another intermediate-age GCs ($6\pm 2$ Gyr) in KK211, which
appears to be the central star cluster of this dSph galaxy. The cluster
age and metallicity estimates are consistent with the brightest
intermediate-age AGB stars in this galaxy, 4$\pm$1 Gyr, $\feh = -1.4 \pm
0.2$, measured by \cite{rejkuba05}.

\subsubsection{GCs in KK221}
The two brightest GCs in KK221 are old ($t>10$ Gyr) and metal-poor
[Z/H]~$\la-1.5$). It is difficult to assess whether these objects are
associated with the center of K221 due to the very faint surface
brightness profile of the host galaxy. However, both seem to be
gravitationally bound judging from their radial velocities.

A mean radial velocity of all GCs in KK221 coincides well with the radial
velocity of the brightest GC. We find a radial velocity anisotropy
(rotation) among the GCs at the 95\%\ confidence level, in the sense that
the radial velocities of GCs 608, 883, and 966 located in the western edge
of KK221 differ systematically from the radial velocities of GCs 1090,
24n, and 27n located in the eastern edge of the galaxy.

\subsection{\afe\ Ratios}
\label{ln:afe}
Because of different progenitor lifetimes type-II and type-Ia supernovae
enrich their ambient medium on different timescales. In consequence the
chemical composition of stellar populations, together with their ages and
metallicities, can be used to constrain their formation timescales. One
good way to do so is to measure the \afe\ ratios of stellar populations.
Massive stars that live up to a few 100 Myr enrich the interstellar medium
predominantly with $\alpha$-elements \citep{ww02}, while type-Ia
supernovae are delayed by $1\!-\!3$ Gyr and eject mainly iron-peak
elements \citep{nomoto97}.

Besides ages and metallicities our fitting routines simultaneously
determine \afe\ ratios for all sample clusters. However, the resolution of
the \afe\ diagnostic grid decreases towards lower metallicities. Most of
our sample GCs have relatively low \zh\ values and the accuracy of \afe\
values is reflected by the range of uncertainties, typically
$\sim\!0.1\!-\!0.3$ dex (see Figure~\ref{diagd} and Table~\ref{ssp}). Our
sample covers a wide range in \afe\ much broader than what is expected
from the average measurement uncertainty and, thus, implies significant
chemical variance in the GC systems of dwarf galaxies. The mean \afe\ for
our sample is consistent with enhanced values with
$\langle$[$\alpha$/Fe]$\rangle=0.19\pm0.04$ for the $\chi^{2}$ technique
and $0.18\pm0.12$ for the iterative approach. This compares well with
values of GCs in the Fornax dSph galaxy in the outskirts of the Milky Way
\citep{strader03a}.

\begin{table*}
\begin{center}
\tabletypesize{\scriptsize}
\caption{Ages, \zh\ and \afe\ for our sample globular clusters determined using our 
$\chi^2$ minimization code and the iterative interpolation routine. The bottom of the table shows the statistics of our sample, i.e. the mean and its error, standard deviation, weighted mean and the corresponding standard deviation, as well as the median and its 25\%-ile margin. See text for details.}
\label{ssp}
\vskip 10pt
\begin{tabular}{lcccrrr}
\hline \hline
Object        & Age$_{\chi^2}$ & \zh$_{\chi^2}$& \afe$_{\chi^2}$ & Age$_{\rm I}$ & \zh$_{\rm I}$ & \afe$_{\rm I}$    \\
	      & (Gyr)      &  (dex)      &  (dex)  & (Gyr)    &  (dex)           &  (dex) \\
\hline
KK211-3-149   & 6$\pm$2    & $-1.4\pm$0.3  &  0.1$\pm$0.3 & $7.4^{3.1}_{1.4}$ & $-1.45^{0.15}_{0.13}$ & $-0.27^{0.20}_{0.23}$\\
KK221-2-966   & 10$\pm$2   & $-1.6\pm$0.1 &  0.1$\pm$0.3 & $11.7^{0.2}_{0.1}$ & $-1.47^{0.04}_{0.02}$ & $ 0.51^{0.08}_{0.08}$\\
KK221-24n     & 9$\pm$2    & $-1.7\pm$0.3 &  0.3$\pm$0.4   & $11.4^{1.1}_{1.0}$ & $-1.58^{0.16}_{0.02}$ & $ 0.28^{0.21}_{0.22}$\\
KK084-3-705   & 9$\pm$1    & $-1.2\pm$0.1 &  0.3$\pm$0.2  & $12.1^{0.2}_{0.2}$ & $-1.16^{0.07}_{0.06}$ & $ 0.99^{0.02}_{0.02}$\\
KK084-3-830   & 10$\pm$4   & $-1.6\pm$0.1 &  0.3$\pm$0.2 & $10.2^{0.9}_{0.1}$ & $-1.45^{0.02}_{0.24}$ & $ 0.25^{0.07}_{0.07}$\\
KK084-4-666   & 8$\pm$3    & $-0.1\pm$0.2 &  0.3$\pm$0.2  & $14.4^{0.6}_{0.6}$ & $-0.31^{0.03}_{0.03}$ & $ 0.52^{0.07}_{0.06}$\\
KK084-36n     & 8$\pm$2    & $-0.5\pm$0.2  &  0.1$\pm$0.2 & $10.7^{1.7}_{1.2}$ & $-0.30^{0.05}_{0.05}$ & $-0.05^{0.05}_{0.04}$\\
U3755-2-652   &  8$\pm$2   & $-1.3\pm$0.2 &  0.1$\pm$0.1 & $10.6^{0.9}_{0.4}$ & $-0.94^{0.03}_{0.02}$ & $ 0.32^{0.09}_{0.09}$\\
U3755-3-914   & 4$\pm$1    & $-1.3\pm$0.1  &  0.0:        & $ 1.1^{0.2}_{0.2}$ & $-0.48^{0.03}_{0.07}$ & $ 0.17^{0.07}_{0.08}$\\
U3755-3-1182  & 2$\pm$1    & $-1.8\pm$0.4 &  0.1$\pm$0.4  & $ 2.3^{0.2}_{1.2}$ & $-2.07^{0.58}_{0.09}$ & $ 0.69^{0.09}_{0.10}$\\
U3755-3-1257  & 7$\pm$2    & $-1.2\pm$0.2 &  0.1$\pm$0.1  & $ 3.2^{0.9}_{0.6}$ & $-0.55^{0.07}_{0.10}$ & $-0.55^{0.08}_{0.08}$\\
U3755-3-2123  & 6$\pm$2    & $-1.8\pm$0.3 &  0.1$\pm$0.4  & $14.0^{0.3}_{2.6}$ & $-2.24^{0.18}_{0.02}$ & $-0.41^{0.06}_{0.02}$\\
U3755-3-2363  & 2$\pm$2    & $-1.3\pm$0.2 &  0.1$\pm$0.1  & $ 0.9^{0.2}_{0.6}$ & $-0.90^{0.31}_{0.07}$ & $-0.11^{0.06}_{0.06}$\\
U3755-3-2459  & 1$\pm$1    & $-1.3\pm$0.1 &  0.4$\pm$0.2  & $ 0.6^{0.1}_{0.1}$ & $-0.36^{0.02}_{0.02}$ & $-0.04^{0.03}_{0.03}$\\
E490-17-2035  & 4$\pm$2    & $-1.4\pm$0.2  &  0.1$\pm$0.1 & $ 7.2^{1.0}_{3.7}$ & $-1.73^{0.89}_{0.29}$ & $-0.34^{0.37}_{0.14}$\\
E490-17-1861  & 9$\pm$4    & $-1.7\pm$0.3  &  0.5:        & $11.7^{0.7}_{1.0}$ & $-1.53^{0.11}_{0.04}$ & $ 0.99^{0.02}_{0.45}$\\ \hline
mean $\pm$ error  & 6.4$\pm$0.7	&$-1.33\pm$0.11	& 0.19$\pm$0.04 		& 8.1$\pm1.2$	&$-1.16\pm0.16$	& 0.18$\pm$0.12 \\
std. deviation $\sigma$          & 3.0                	& 0.46                   	& 0.14		& 4.9			& 0.63			& 0.47 \\
median$\pm$25\%ile
			  & 8.0$\pm$2.5	& $-1.3\pm0.2$	& 0.1$\pm$0.1
		& 10.6$\pm$4.3	& $-1.16\pm0.53$ & 0.25$\pm0.32$ \\
\hline \hline
\end{tabular}
\end{center}
\end{table*}

Interestingly, the mean \afe\ of GCs in UGC~3755 is systematically lower
than that of the rest of the sample. For the mean \afe\ of UGC~3755 we
find $\langle$[$\alpha$/Fe]$\rangle=0.13\pm0.05$ using the $\chi^{2}$
technique and $0.01\pm0.16$ with the iterative approach, while the rest of
the sample has $\langle$[$\alpha$/Fe]$\rangle=0.23\pm0.05$ using the
$\chi^{2}$ technique and $0.32\pm0.16$ using the iterative approach. In
each case, this is a $\sim\!2\sigma$ offset. In particular, the youngest
GCs in UGC~3755 have the lowest \afe\ ratios, but the small number
statistics of this sub-sample makes the age-\afe\ correlation marginally
significant and we do not attempt to quantify this trend.

The two most metal-rich GCs, KK84-666 and KK84-36n, are an interesting
pair in terms of their \afe\ ratios. Although KK84-666 has roughly solar
metallicity (see Table~\ref{ssp}) its $\alpha$-element enhancement is
relatively high, and comparable to those of Local Group GCs. KK84-36n, on
the other hand, which has the second highest metallicity in our sample
shows almost no $\alpha$-element enhancement, and clearly stands out
compared with the rest of GCs in KK84. Together with its location, which
puts it closer to NGC~3115 than KK84 (in terms of projected distance), its
low \afe\ ratio is yet another piece of evidence that this GCs is likely a
member of the extended disk of NGC~3115 (see also Sect.~\ref{agekk84}).

We observe a relatively large fraction of low-\afe\ GCs which are less
frequent in the GC systems of Local Group spirals. In our sample
$44\pm17\%$ of GCs have sub-solar \afe\ ratios. In comparison, only
$\sim20\!-\!30$\% of GCs in the two Local Group spirals Milky Way and M31
which had their \afe\ determined with the same technique have sub-solar
\afe\ values \citep{p06}. We point out that the sampled age, metallicity,
and luminosity ranges for GCs in nearby LSB dwarf galaxies and Local Group
spirals are not the same and that correlations between the \afe\ and any
of these parameters will likely alter the observed fraction of sub-solar
\afe\ clusters. If we restrict the Local Group spiral GC sample to
intermediate and faint luminosity ($M_B \ga - 8$), old ($t > 8$ Gyr), and
metal-poor GCs ([Z/H]~$<-1$) - the typical regime of our observed sample -
the fraction of sub-solar \afe\ clusters drops to $\sim\!15\%$, which is
about 1.7 $\sigma$ off compared to the fraction in our sample. The work of
\cite{pritzl05} which derives mean \afe\ ratios for Milky Way GCs from
high-resolution spectroscopy of individual member stars shows an even
lower fraction of $\sim\!2$\%. Although this is the most representative
comparison we can perform at the current state (given the limited sample
statistics), we stress that GCs in dwarf and spiral galaxies are likely to
have experienced different dynamical evolution histories. Star clusters of
similar luminosity (mass) we observe today may have started off with very
different initial masses. However, since only the most massive GCs are
prone to self-enrichment \citep[such as $\omega$Cen, e.g.][]{recchi05,
villanova07}, the GCs in our sample should reflect the global chemical
composition at the time of their formation, independent of their
mass.~Hence, we speculate that the somewhat lower fraction of Milky Way
GCs with sub-solar \afe\ may be an indication for significantly shorter
star-formation and enrichment timescales compared to those in field dwarf
galaxies.


\subsection{Other Abundances}

\begin{figure*}[!ht]
\centering
\includegraphics[width=7.7cm]{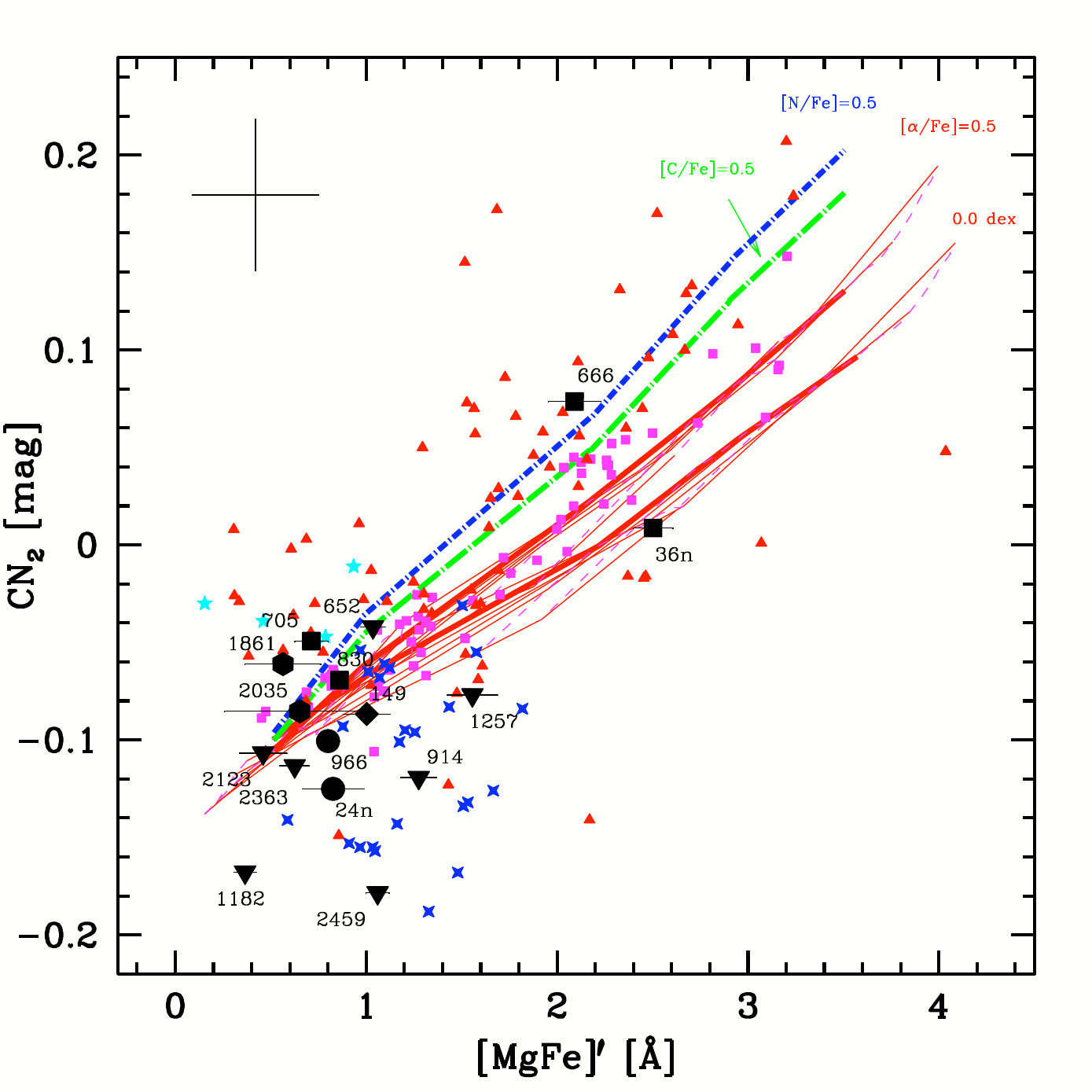}
\includegraphics[width=7.7cm]{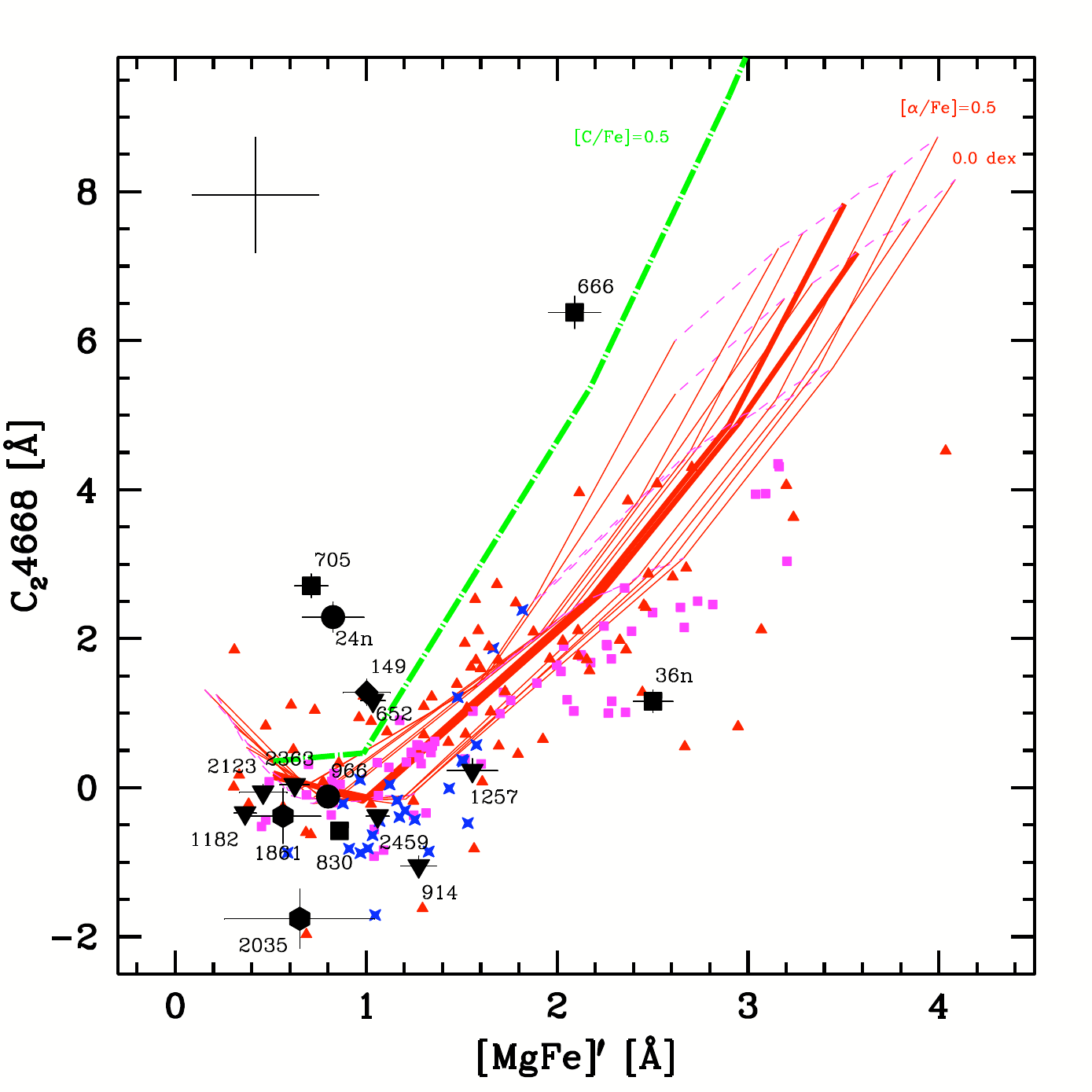}
\includegraphics[width=7.7cm]{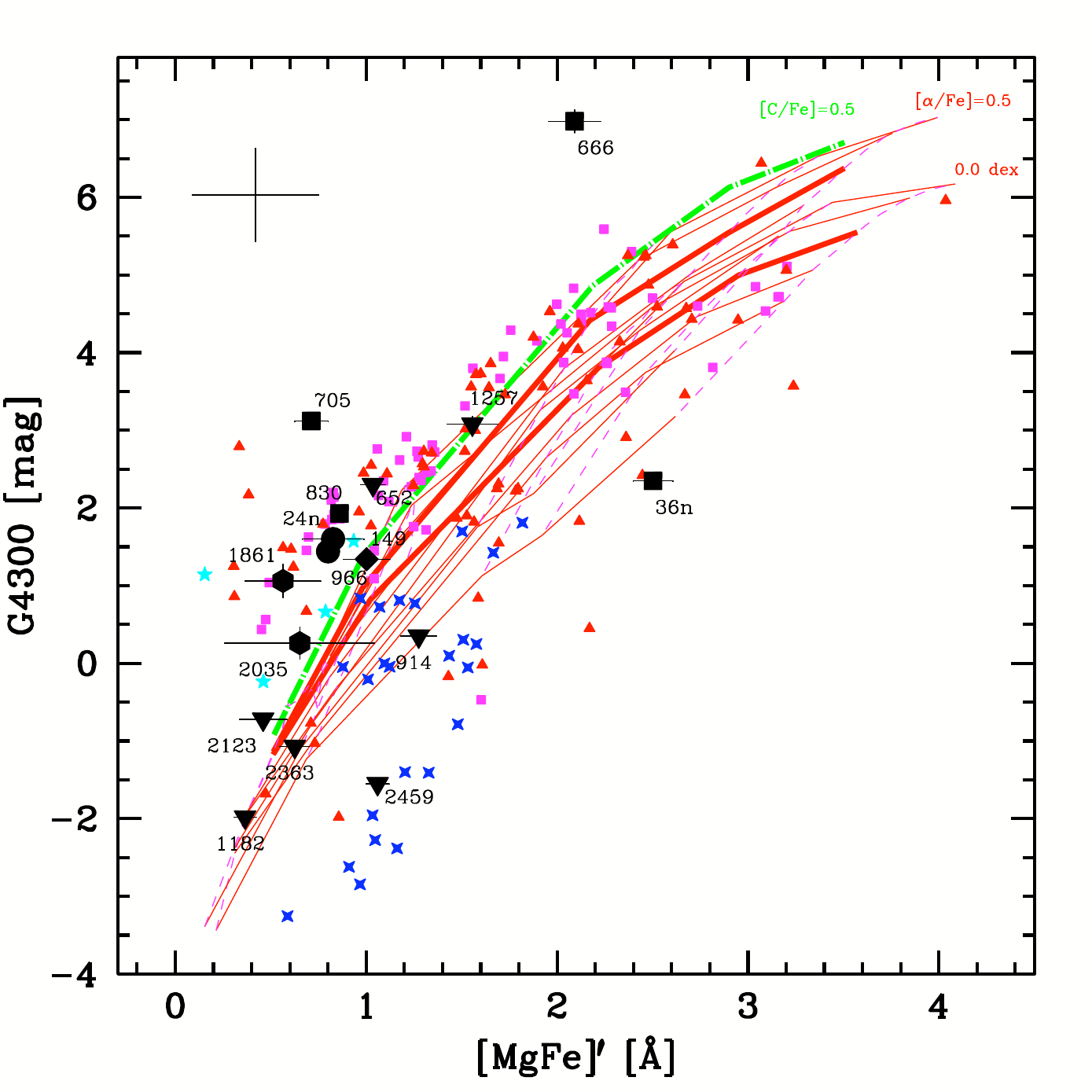}
\includegraphics[width=7.7cm]{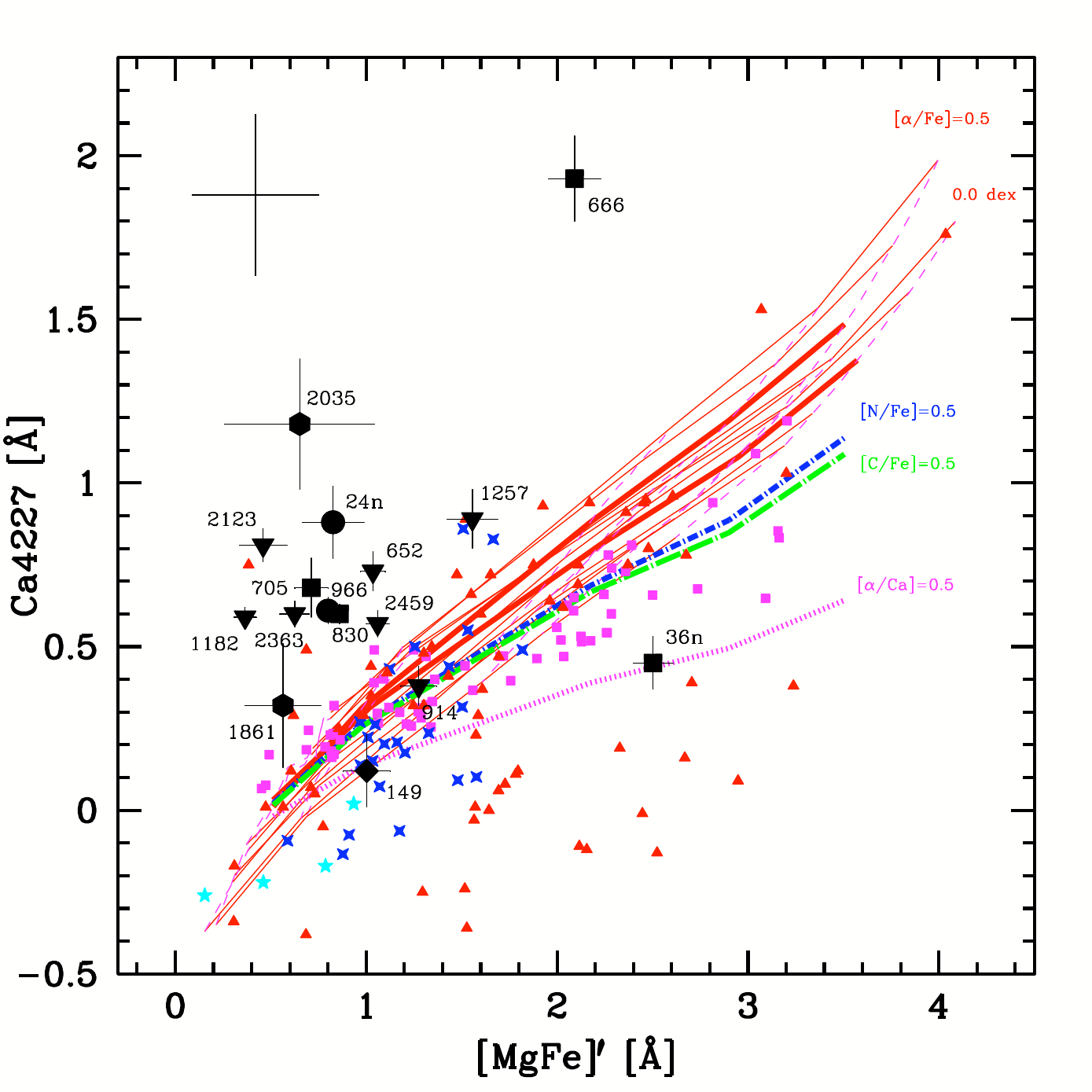}
\caption{Diagnostic plot of [MgFe]\arcmin\ versus indices primarily sensitive 
to C, N, and Ca abundances. We plot two model grid for \afe=0.0 and 0.5 dex, 
and all ages and metallicities as in Figure~\ref{diagd}. Note the offset in CN
between younger UGC~3755 GCs and other GCs.}
\label{cn_mgfe}
\end{figure*}

We conducted a detailed investigation of numerous other Lick index
diagnostic diagrams. In Figure~\ref{cn_mgfe} we show diagnostic plots for
indices that are sensitive to the abundance of carbon, nitrogen, and
calcium \citep{tripicco95}. Because a strict quantitative analysis is not
possible due to a lack of corresponding model predictions, we focus the
following discussion on qualitative trends.

\subsubsection{Carbon and Nitrogen}
The CN$_2$ versus [MgFe]\arcmin\ diagram (upper left panel) shows
significant scatter in CN index strength for our sample GCs, mostly
"below" the model grid towards lower CN$_2$ index values. At metallicities
typical for our sample GCs, this scatter cannot be accounted for by \afe\
variations alone as it is demonstrated in the Figure by the two grid of
population synthesis models for \afe\ ratios of 0.0 and $+0.5$ dex. It
requires additional variance in, at least, one other element abundance to
which the CN$_2$ index is sensitive to match the observed distribution. To
qualitatively test this hypothesis we explore the influence of carbon and
nitrogen enhancement on the model grid using the predictions for C and
N-enhanced models from \cite{thomas03}. We overplot two isochrones for
factor 3 enhanced C and N abundance at \afe$=0.5$ for a 5 Gyr old stellar
population. We recall that the corresponding predictions for non-enhanced
populations are plotted as thick lines in the two grids. The shape of the
C/N-enhanced model grid is very similar to that of the non-enhanced
predictions and for clarity reasons we avoid plotting the entire grid for
the C/N-enhanced populations, and note that age and metallicity are highly
degenerate in these grids and that both parameters have no impact on the
following discussion.

The comparison of the enhanced models with our data in the CN$_2$
vs.~[MgFe]\arcmin\ diagram shows that some GCs in dwarf galaxies appear to
be highly under-abundant in C and/or N. This is particularly the case for
GCs in UGC~3755, which exhibit on average significantly lower CN$_2$ index
values. However, from the variations in CN$_2$ index strength alone we
cannot decide whether chemical variance in carbon and/or nitrogen is
responsible for the offsets.

A sanity check is provided by the C$_2$4668 and G4300 vs.~[MgFe]\arcmin\
diagrams which are both mildly sensitive to C abundance but not sensitive
to N variations \citep{tripicco95}. Most GCs in the C$_2$4668 and G4300
vs. [MgFe]\arcmin\ diagrams show relatively little deviations from the
general trend of model predictions, and we infer that the abundance of
carbon is less likely to change with respect to predictions of standard
population synthesis models than the nitrogen abundance, which implies
that the nitrogen abundance appears to vary significantly in GC systems of
dwarf galaxies. We suggest that this pure qualitative result is being
confirmed with higher resolution spectroscopic observations.

\subsubsection{Calcium}
In the Ca4227 vs. [MgFe]\arcmin\ diagram we add a model which describes a
stellar population with a factor 3 calcium under-abundance relative to the
other $\alpha$-elements. In the Ca4227 vs. [MgFe]\arcmin\ plot (lower
right panel in Figure~\ref{cn_mgfe}) many GCs in our sample show excess in
Ca4227 index strength. The overplotted model for [$\alpha$/Ca]~$=0.5$
indicates that this offset may be due to a Ca enhancement in some GCs.~As
the absolute calibration of Ca abundance model predictions is still
uncertain \citep[see][]{cenarro04, prochaska05}, we merely point out this
rather intriguing abundance pattern should be checked with higher
resolution spectra.


\section{Discussion}
\label{discussion}

\subsection{Chemical Tagging of GCs in LSB Dwarf Galaxies}

The variance of GC chemical compositions in and outside the Local Group
reveals complex enrichment histories.~The {\it old} GCs in our sample
with ages $t\ga8$ Gyr show marginally different \afe\
ratios\footnote{$\langle$[$\alpha$/Fe]$\rangle=0.21\pm0.04$ using the
$\chi^{2}$ technique and $0.25\pm0.15$ dex for the iterative approach.}
compared to the typical old GCs in the Local Group, which are
$\alpha$-enhanced at $0.29\pm0.01$ dex \citep[leaving out Pal12, Ter7,
Rup106, and M68, see][for details]{pritzl05}. The younger GCs in our
sample with ages $t\la8$ Gyr have significantly lower \afe\
ratios\footnote{$\langle$[$\alpha$/Fe]$rangle=0.14\pm0.07$ using the
$\chi^{2}$ technique and $0.03\pm0.20$ dex for the iterative approach.}
and are comparable to Milky Way GCs associated with the Sagittarius
remnant and other distinct GCs such as Ruprecht 106, $\omega$ Centauri,
NCG~2419, which have $\langle$[$\alpha$/Fe]$\rangle=0.06\pm0.05$ 
\citep{pritzl05}.

Solar-type \afe\ ratios are consistent with star formation timescales
longer than $\sim\!1\!-\!3$ Gyr, when ejecta of type-II {\it and} type-Ia
supernovae are fully mixed in the interstellar medium \citep{greggio05}.
\afe\ ratios at $\sim\!0.3$ dex indicate shorter and more intense
cluster formation, on timescales of the order of a few hundred million
years. This, in turn, suggest that some of the oldest GCs in our sample
were formed relatively early, at similar epochs as the typical Milky Way
GC. Less $\alpha$-enhanced GCs, on the other hand, likely formed $\ga\!1$
Gyr after the Big Bang at $z\!\approx\!5.7$ or later, which would place
their formation period at the end of reionization or thereafter
\citep[e.g.][]{kashikawa06, benson06}.

The characteristic chemical compositions limit the fraction of accreted
GCs from satellite LSB-type galaxies during the assembly process of Local
Group spirals galaxies. Given the difference in $\alpha$-enhanced to
non-$\alpha$-enhanced GCs between our sample and the Local Group spirals
(see Sect.~\ref{ln:afe}), our results imply that to qualify as potential
building block for the two massive Local Group spirals in the hierarchical
picture of galaxy formation, LSB dwarf galaxies would either have to {\it
i)} cease forming star clusters long before the beginning of enrichment by
type-Ia supernovae before being accreted much later by a more massive halo
or {\it ii)} being accreted when the gas out of which star clusters were
forming was still not polluted by type-Ia supernovae.

Other element abundance ratios provide us with the opportunity to
chemically tag GCs that were formed in the field environment and later
accreted by more massive galaxy halos, much like the chemical tagging of
accreted stellar sub-populations that are part of the diffuse-light
component in nearby galaxies \citep{fh02, geisler07}. For instance, at a
given [MgFe]\arcmin\ index (i.e. total metallicity) the carbon and
nitrogen enhancement of our sample GCs is consistent with that of Galactic
GCs, but fails to match the chemical composition of M31 GCs which have
excess CN$_2$ indices (see Fig.~\ref{cn_mgfe}).~It was shown in a series
of studies that the higher CN$_2$ index values for old M31 GCs are due to
a nitrogen enhancement by at least a factor of three compared to the
younger cluster population \citep{li03, burstein04, beasley05, p05a}.~We
remind the reader that the comparison sample of M31 GCs is biased towards
disk clusters and samples only poorly the halo GC population \citep[see
Fig.~1 in][]{p05a}, where accreted objects are more likely to
reside.~Interestingly, although the CN$_2$ indices of the N-enriched M31
GCs do not match our sample GCs, similar N-enhancement can be suspected in
GCs of the Fornax dSph galaxy (see Fig.~\ref{cn_mgfe}).~We conclude that
it is relatively unlikely that M31 accreted a significant amount GCs from
satellite dwarfs similar to those in our sample on orbits close to the
plane of the M31 disk. On the other hand, judging from Figure~\ref{cn_mgfe} 
the average C/Fe and N/Fe chemistry of our sample
GCs appears to be similar to the one of metal-poor Milky Way GCs. We currently
lack the spectroscopic database of outer-halo GCs in the two Local Group
spirals to further constrain the selective accretion history of these
massive spiral galaxies.


\subsection{Specific Frequencies of GCs in LSB Dwarf Galaxies}

\begin{figure}
\centering
\includegraphics[width=6cm,angle=-90,bb=60 60 567 732]{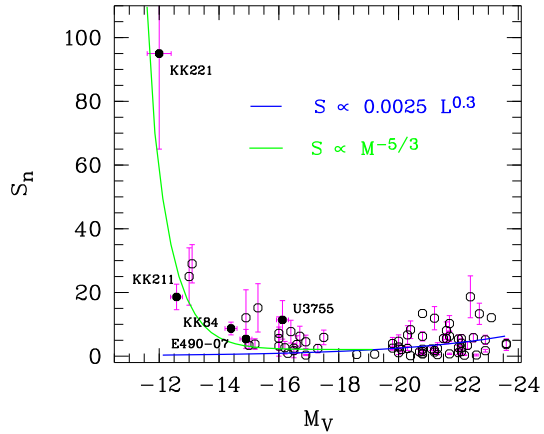}
\caption{Specific frequency, $S_{N}$, versus luminosity for
galaxies from the samples of \cite{durrell96} and \cite{harris91} ({\it
open symbols}), and for our sample of five galaxies ({\it solid circles}).
The dark solid line shows $S_{N}$ predictions which assume that GCs were
formed in galaxies in direct proportion to the initial gas mass. The light
solid line shows the Dekel and Silk model of mass loss.}
\label{ushape}
\end{figure}

The specific frequency, $S_N$, (Harris \& van den Bergh 1981) is defined
as the number of globular clusters per unit $M_V\!=\!-15$ mag of host
galaxy light: $S_{N}\!=\!N_{\rm GC} \cdot 10^{0.4(M_V+15)}$.~It is tightly
related to cluster formation efficiency, which is varying as a function of
host galaxy's morphological type, mass, and local environmental density
\citep[e.g.][]{harris91, richtler95}. \cite{kumai93} and \cite{west93}
found a significant trend of increasing $S_{N}$ with increasing
environmental density. Another observational fact is that $S_{N}$ values
grow for nucleated dwarf galaxies with decreasing galactic mass
\citep[e.g.][]{miller98}.

Figure~\ref{ushape} illustrates the $M_V - S_{N}$ diagram, where the
galaxy luminosity is plotted against the GC specific frequency. Our five
sample LSB galaxies are shown as solid circles, while data for dwarf
galaxies in the Virgo cluster \citep{durrell96} and more massive
early-type galaxies \citep{harris91} are shown as open symbols. The solid
line is a model for which $S_{N} \sim 0.0025\, L_{\rm gal}^{0.3}$
\citep{mclaughlin99}, where a constant number of GCs per unit mass is
formed with an efficiency $e=0.0025$, and the mass of gas is much lower
than the stellar mass. This thin line shows the mass-loss model $S_{N}
\sim M^{-5/3}$, which follows from $ M/L_V \sim M^{2/3}$ \citep{dekel03}
or $M/L \sim L^{-0.37}$ \citep{dekel86}. The vertical normalization of
this model is arbitrary and was chosen to fit the dwarf galaxy data.

We point out the similarity of $S_{N}$ values for field and cluster dwarf
galaxies at a given galaxy luminosity, especially for faint galaxies.
Provided not a size-of-sample effect, this result implies that environment
is not the driving parameter for the $M_V - S_{N}$ correlation for dwarf
galaxies. Less massive galaxies lose gas more efficiently because objects
with shallower potential wells develop galactic winds more easily (e.g.,
Arimoto \& Yoshii 1987; Matteucci 1994), and it seems plausible that
internal factors, such as galactic winds, are shaping the $M_V - S_{N}$
relation of dwarf galaxies. These results require further testing with
larger galaxy samples.

A particularly interesting fact is that dIrr and dSph galaxies seem to
follow the same $M_V - S_{N}$ trend. At face value, this appears to be in
contrast with chemical evolution models and observations of abundance
ratios in low-mass galaxies. For instance, \cite{lanfranchi03} concluded
that one or two long starbursts with very efficient winds well describe
the chemical evolution of dSphs. On the other hand, blue compact galaxies
are characterized by a star formation history proceeding in several short
bursts separated by long quiescent periods. Given varying star formation
histories, similar $M_V - S_{N}$ relations for both dIrr and dSph galaxies
require that the fading of the galaxy light and the dynamical evolution of
the globular cluster system are tightly related. 

Passive evolution of dIrr galaxies leads to a fading of up to $\sim\!2$
mag of their integrated magnitudes if they were to abruptly stop forming
stars \citep{hunter85}. Assuming a non-changing $S_{N}$, this requires the
disruption of $\sim\!80\!-\!90\%$ of all formed star clusters, and is in
line with observations of the so-called cluster ``infant mortality'' in
nearby young star cluster systems \citep{chandar06, whitmore07}. This also
suggests that the effects of tidal forces acting in group and cluster
environments are linked in disrupting star clusters and their host
dIrr LSB galaxies \citep{gnedin03, georgiev06}. We suggest a detailed
study of globular cluster systems in low-mass galaxies sampling galaxy
mass, morphology, and environmental density to test our findings.


\subsection{Nuclei of Dwarf Galaxies}
Our spectroscopic study shows that some of the lowest mass galaxies ($M_V
\approx -12$ mag) can have nuclear star clusters, i.e.~the brightest GC is
located near the optical center of a galaxy. The two most prominent cases
in our data set are KK211-149, and KK84-830. Both have low metallicities
[Z/H]~$\approx-1.5$, see Tab.~\ref{ssp}). KK211-149 is an intermediate-age
GC, while KK84-830 is old. Both nuclear clusters are found in dwarf
galaxies located in close vicinity to massive galaxies, NGC~5128 and
NGC~3115, respectively.

\cite{mclaughlin06} have shown that the limiting mass of a central massive
object in dwarf galaxies is defined as follows:
\begin{equation} M_{\rm CMO} = 3.67\times 10^8 \ M_\odot\ \lambda^{-1}\,
\sigma_{200}^4\, (f_g/0.16)\ , \end{equation} where $\lambda$ is the
massive-star feedback equal to $\sim\!0.03-0.1$ for a nuclear star
cluster, $\sigma_{200}\equiv \sigma/200\ {\rm km\,s}^{-1}$, and the baryon
fraction $f_g = \Omega_b/\Omega_m = 0.16$ \citep{SpergelEtal03}. So, a
limiting mass of a nucleus in dSph galaxies with $ \sigma \approx 10$
$km\,s^{-1}$ is in the range $\sim\!10^4\! -\! 10^5 M_\odot$, the mass
range of Local Group globular clusters. We compute the total masses for
both our nuclear clusters from the photometric information in
Tables~\ref{gcprop} and \ref{ssp} using the population synthesis
prediction of \cite{bc03} assuming a Salpeter IMF. We find $2.5\times10^5
M_\odot$ for KK211-149 and $2\times10^6  M_\odot$ for KK84-830. A more
"top-heavy" IMF compared to the Salpeter IMF would bring both mass
estimates in better agreement. Indications for a "top-heavy" IMF are found
in the central stellar populations of the Milky Way \citep{nayakshin06,
nayakshin07}. Heated molecular clouds are suspected to produce "top-heavy"
IMFs due to a simultaneous increase in the thermal Jeans mass and the
collisional destruction of low-mass stellar cores \citep{elmegreen03}.

The two nuclear star clusters are massive and compact enough to survive a
Hubble time in isolated galaxies in the absence of dynamical factors, such
as tidal interactions, galaxy-galaxy encounters, interaction with ISM,
etc.~Using the structural parameters determined in \cite{sh05} and the
evolution models of \cite{fz01} we compute a mass-loss of 7\% for
KK211-149 and 3\% for KK84-830 over the next 12 Gyr due to two-body
relaxation. Including tidal interactions on orbits typical for Milky Way
GCs, as adopted by \citeauthor{fz01}, these fractions increase by a factor
$\la\!2$. So far we know of only one nucleated early-type dwarf galaxy
(Sagittarius dSph) in the Milky Way subgroup, a few in the M31 subgroup,
and none in the Canes Venatici Cloud, and among isolated nearby LV
galaxies \citep[e.g.][and references therein]{grebel06}. The absolute
number of nucleated early-type dwarf galaxies is higher in denser
environments, such as the Fornax and Virgo galaxy clusters
\citep[e.g.][]{bin85, miller98, cote06, lisker07}. Although this suggests
that the process of nucleation in cluster dwarf galaxies is likely driven
by dynamical factors that depend on the local environmental density, the
nucleation fraction among dwarf galaxies in different environments appears
to be roughly constant. Compared to the few nucleated dwarfs in the Local
Group with a total mass of ${\cal M}_{\rm total}\approx2\cdot10^{12}
M_{\odot}$ \citep{kara05, vdb06}, the Virgo galaxy cluster holds
$\sim\!300$ nucleated dwarfs \citep{sandage85} and has a total dynamical
mass of ${\cal M}_{\rm total}\approx1.2\cdot10^{15} M_{\odot}$
\citep{fouque01}. Hence, our back of the envelope calculation suggests
similar nucleation fractions as a function of the total group/cluster
mass. But, of course, these numbers are very rough, especially for the
Virgo cluster, where the number of galaxies with nuclear star clusters
grows with higher spatial resolution \citep{cote06}. Clearly, a larger
sample is necessary to assess the frequency of nuclear clusters among
field dwarf galaxies.


\section{Conclusions}
\label{conclusion}

Numerous photometric and spectroscopic studies of globular clusters in
Virgo and Fornax cluster dwarf galaxies have been undertaken in the last
years, which targeted bright dwarf galaxies down to $M_V\approx-15$ mag
\citep[see][]{miller06}. Due to observational selection effects dwarf
galaxies fainter than this are missed at distances of $D \approx 17$ Mpc.
Faint LSB dwarf galaxies down to $M_V \approx -12$ mag have long been
thought to be free of globular clusters, because they have insufficient
mass. Our HST/WFPC2 survey of low-mass dwarf galaxies (SPM05), situated at
distances $2-6$ Mpc in the Local Volume, revealed a rich population of
globular cluster candidates (GCCs). In this work, we observed five of
these galaxies with the VLT/FORS2 spectrograph in MXU mode and found that
all targeted GCCs except one are genuine globular clusters. We could also
confirm five additional globular clusters in our sample galaxies. Two
clusters appear to be the nuclei of KK84 and KK211. The confirmed globular
clusters are in general old and metal-poor, and show a range of \afe\
ratios. The mean $\langle$[$\alpha$/Fe]$\rangle=0.19\pm0.04$ that was
determined with the $\chi^{2}$ minimization technique and $0.18\pm0.12$
dex which was computed using the iterative approach appears slightly lower
than the mean $\langle$[$\alpha$/Fe]$\rangle=0.29\pm0.01$ for typical
Milky Way clusters. Globular clusters in the two isolated, relatively
bright dwarf galaxies UGC~3755 and ESO~490-17 show a wide range of ages
from 1 to 9 Gyr, and imply extended star formation histories in these
galaxies. This goes in hand with the measured low \afe\ ratios for the
younger clusters and is consistent with low intensity star bursts. The
oldest clusters with the highest \afe\ are found in KK84, a companion of
NGC~3115. Other chemical abundances indicate potentially interesting
differences between globular clusters in dwarf and more massive galaxies
and, if confirmed, would facilitate the quantification of the accreted
mass in rich GC systems of massive early-type galaxies.

\acknowledgments

We thank the referee for a constructive report that helped to improve the
paper. THP gratefully acknowledges support in form of a Plaskett
Fellowship at the Herzberg Institute of Astrophysics. MES thanks
D.I.~Makarov for his help with programming in Matlab. We are grateful to
Ricardo Schiavon for providing his spectroscopic Milky Way GC data in
electronic form. This work is based on observations made with ESO
Telescopes at the Paranal Observatory under program ID P76.A-0137 and
P76.B-0137.

\newpage
\appendix

\section{Surface Brightness Profiles}
\label{sbprof}

\begin{figure*}[!ht]
\centering
\includegraphics[width=11cm]{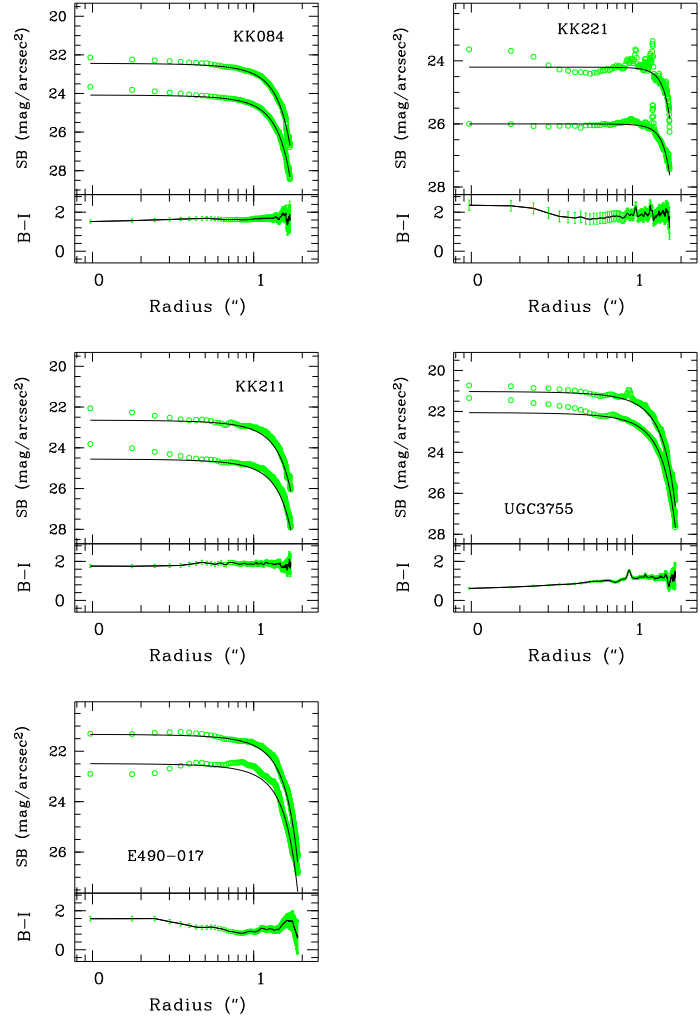}
\caption{Azimuthally averaged surface brightness profiles and $B\!-\!I$ color
profiles for our sample dwarf galaxies. In all panels, the upper surface
brightness curve belongs to the $B$-band measurement.}
\label{surf}
\end{figure*}

Fundamental structural parameters of galaxies determined from their
surface brightness (SB) profiles provide valuable information about
processes of galaxy formation \citep[e.g][]{kormendy85, dekel86}. In
Tables~\ref{dwgprop} and \ref{dwgprop1} we show fundamental photometric
parameters obtained for our sample galaxies using our SB profiles. The $B$
and $I$ SB profiles and the corresponding $B-I$ distributions are
presented in Figure~\ref{surf}. The errors of SB profile determination
depend primarily on the accuracy of the background estimates and the
position of the galactic center. The background estimates are relatively
uncertain in cases where bright stars are projected on top of the SB
profile. That is why our surface photometry results should be taken as
rough estimates for this extremely low surface brightness dwarf galaxy
KK221. The choice of a center is complicated for dwarf irregular galaxies
with multiple, bright star-forming regions. Given all uncertainties we
estimate typical error of the integrated visual magnitude of the order of
$\sim\!0.2$ mag. The errors of the SB profiles are oveplotted in the
distributions of colors in Figure~\ref{surf}.

The solid lines overplotted on the SB profiles show the Sersic law
approximation. The Sersic model \citep{sersic68} describes SB profiles of
the form: $$ I(r)=I_0 \exp[-\nu_n(r/r_e)^n] $$ where $I(r)$ is the SB (in
intensity) at radius $r$, $I_0$ is the central surface brightness, $r_e$
is the effective half-intensity radius, $\nu_n \simeq 2n - 1/3
+4/(405n)+46/(25515n^2)$, and $n$ is the dimensionless shape parameter
that determines the curvature of the profile. For $1/n > 1$ the profiles
become flat in the central part, while for $1/n< 1$ they are cuspy. The
best-fitting parameters of the Sersic profile are presented in
Table~\ref{dwgprop1}. It should be noted that central bright objects such
as GCs or star forming regions were excluded from the fit.

Table~\ref{dwgprop2} summarizes the literature data on the fundamental
photometric parameters of our sample galaxies. One can see, that in
general our photometric results agree fairly well with the literature
data. However, the shape parameter $n$ determined by \cite{jerjen00} for
KK211 implies a flatter profile than ours. The color profile is irregular
for our sample dIrrs and for KK221, for which the sky subtraction is
difficult.

The Sersic index $n$ appears to be similar for all our sample dwarf
galaxies, except KK221, which may be tidally disrupted. The SB profile for
KK221 is two times flatter than for the other sample galaxies. Similarity
of fundamental parameters for rotating and non-rotating dwarf galaxies of
different morphological type and situated in different environments may
indicate that the internal structure of such faint galaxies is primarily
defined by their mass.

\section{Mass estimate for KK221}

To derive the mass of KK221 we use the mass estimator for tracer
populations \citep{evans03}:
\begin{equation}
M_{\rm press}=\frac{C}{GN}\sum_i (v_{i, {\rm los}}-\langle v\rangle)^2 R_i,
\end{equation}
where
\begin{equation}
C=\frac{16(\gamma-2\beta)}{\pi
(4-3\beta)}\cdot\frac{4-\gamma}{3-\gamma} \cdot\frac{1-(r_{\rm
in}/r_{\rm out})^{3-\gamma}}{1-(r_{\rm in}/r_{\rm out})^{4-\gamma}}.
\end{equation}
Here, $\langle v\rangle$ is the system's mean radial velocity and $\beta$
the anisotropy parameter $1-\sigma_t^2/\sigma_r^2$ which is unity for
purely radial orbits and $-\infty$ for a system with solely tangential
orbits \citep{binney81}.

Using the power-law rule $\gamma 1+d\log\Sigma / d\log R$
\citep{gebhardt96} to derive the three-dimensional density profile of the
GC population, we obtain $\gamma = 1.4$ from the Sersic index $n$ of KK221
SB distribution (see Table~\ref{dwgprop}), and $M_{\rm press} \approx 3
\cdot 10^8 M_{\sun}$. We use here the anisotropy parameter $\beta=0.5$ for
randomly oriented orbits, and radii of orbits of the nearest and most
distant GC $r_{\rm in}=0.9$ kpc and $r_{\rm out}=1.8$ kpc \citep{sh05},
correspondingly. The statistically unbiased estimate of the mass value is
$M_{\rm press}^c= M_{\rm press}[1-(2 \sigma_v^2)/3 s_v^2]$, where
$\sigma_v$ is the rms error of the radial velocity measurements, and $s_v$
is the rms velocity of GCs relative to the mean velocity of GC system,
$s_v^2= (1/k)\sum{(v_k-\langle v \rangle)^2}$ with a number of GCs equal k
\citep{kara99}. $M_{\rm press}^c= 0.2 M_{\rm press}$ in our case. So, we
obtain the statistically unbiased estimate of the mass for KK221 $M_{\rm
press}^c \sim 6 \cdot 10^7 M_{\sun}$, and the corresponding mass-to-light
ratio $M/L_v \approx 15$. Corrected for the effect of radial velocity
anisotropy (Sec.~\ref{evol_par_GCs}), the mass-to-light ratio $M/L_v$
appears to be $\sim 9 M/L_{v, \sun}$.~The expected uncertainty of the
total mass estimate is $\sim\!50$\% taking into account the small number
of GCs. In our case where the number of GCs is six, the error of the
velocity dispersion measurement is $\sim\!33$\% of the value of the
velocity dispersion itself.

\begin{table}
\begin{center}
\scriptsize
\caption{Fundamental photometric parameters for our sample galaxies from literature
sources.}
\label{dwgprop2}
\vskip 10pt
\begin{tabular}{lllccll}
\hline \hline
Galaxy      &  B       & $B-R$         & $R_e(B)$ & $\mu_0(B)$ & h(\arcsec) & Reference   \\
\hline
KK211       & 16.32    & 1.56          & 21.1    & 24.48     & $1.72^{(1/n)}$ & \cite{jerjen00}  \\
KK84        & 16.16    & 1.38          & 17.99   & 23.19     & 10.59      & \cite{parodi02}  \\
U3755       & 14.07    & $0.55^{(B-V)}$   & \nodata & 21.64     & 12.62      & \cite{mak99}  \\
E490-017    & 13.67    & 0.83          & 22.83   & 21.30     & 13.28      & \cite{parodi02}  \\
\hline \hline
\end{tabular}
\end{center}
\tablecomments{ Columns contain the following data:
(2), (3) total B magnitude and ($B\!-\!R$) color (for UGC3755 ($B\!-\!V$) is given),
(4) effective radius, (5) central surface brightness, (6) exponential
scale length (for KK211 the Sersic profile shape parameter is given), (7)
reference.}
\end{table}

\begin{table}[!ht]
\begin{center}
\scriptsize
\caption{Fundamental parameters of our sample dwarf galaxies derived from the
surface photometry on the VLT/FORS2 images.}
\label{dwgprop1}
\vskip 10pt
\begin{tabular}{lrcccll}
\hline \hline
Galaxy  &  $\mu_0(B)$    &  $\mu_0(I)$   & $R_e(B)(\arcsec)$&      n      &   $R_c(\arcsec)$        &  $R_t(\arcsec)$    \\
\hline
KK221   & 26.00$\pm$0.20 & 24.20$\pm$0.20 & 32.3$\pm$0.5  & 0.40$\pm$0.20 & 78$\pm$19    & 85$\pm$9   \\
KK211   & 24.51$\pm$0.12 & 22.60$\pm$0.15 & 24.1$\pm$0.1  & 0.85$\pm$0.15 & 24$\pm$1   & 89$\pm$1  \\
KK084   & 24.04$\pm$0.07 & 22.40$\pm$0.07 & 19.4$\pm$0.4  & 0.79$\pm$0.01 & 17$\pm$1   & 75$\pm$1    \\
U3755   & 22.01$\pm$0.05 & 20.98$\pm$0.10 & 22.6$\pm$0.1  & 0.85$\pm$0.05 & \nodata        & \nodata     \\
E490-17 & 22.45$\pm$0.20 & 21.29$\pm$0.05 & 26.4$\pm$1.2  & 0.76$\pm$0.04 & \nodata        & \nodata     \\
\hline \hline
\end{tabular}
\end{center}
\tablecomments{Columns contain the following data:
(2) central surface brightness in B band with the corresponding error (a
mean value between the corresponding best-fitting parameters of Sersic and
King profiles), (3), (4) best-fitting parameters of the Sersic profile,
effective radius and degree with corresponding errors, (5), (6) the King
law approximation parameters, core radius and effective radius with
corresponding errors.}
\end{table}


\section{Lick Index Measurements}

We provide the calibrated Lick index measurements of all confirmed GCs in Table~\ref{lickind1}.

\begin{table}[!h]
\begin{center}
\caption{Globular cluster indices (first line) corrected for zeropoints of transformation to the standard Lick system and errors determined from bootstrapping of the object spectrum (second line).}
\label{lickind1}
\tiny
\begin{tabular}{lrrrrrrrrrrrrrrrrrrrrrr} \\
\hline \hline
ID                  &H$\delta_{\rm A}$ &H$\gamma_{\rm A}$& H$\delta_{\rm F}$ &H$\gamma_{\rm F}$&CN$_1$  & CN$_2$ & Ca4227 & G4300 & Fe4383 & Ca4455  & Fe4531 & Fe4668 & H$\beta$ & Fe5015 & Mg1  & Mg2       & Mgb   & Fe5270 & Fe5335 & Fe5406 \\
(S/N)                &(\AA)             & (\AA)          &  (\AA)             &     (\AA)       &(mag)   & (mag)  & (\AA)  & (\AA) & (\AA)  & (\AA)   &  (\AA)& (\AA)  &  (\AA)   & (\AA)  & (mag)& (mag)     & (\AA) & (\AA)  & (\AA)  & (\AA)  \\ \hline
		     &                  &                &                    &                 &        &        &        &       &        &         &       &        &          &       &        &          &        &        &       &        \\
{\bf KK211}          &                  &                &                    &                 &        &        &        &       &        &         &       &        &          &       &        &          &        &        &       &        \\
{\bf 149}            & 4.19             & 1.41           &   2.98             & 2.69            &-0.1272 & -0.0867 &  0.12 &  1.34 &   1.93 &   0.07  & 1.19 &   1.28 &   3.33   & 2.56  & -0.0142 &  0.0538 &  1.05  &  0.93  &  1.03 &   0.21  \\
(40) \hskip 9pt $\pm$& 0.22             & 0.23           &   0.23             & 0.23            & 0.0021 &  0.0031 &  0.11 &  0.12 &   0.14 &   0.14  & 0.15 &   0.18 &   0.18   & 0.19  &  0.0050 &  0.0050 &  0.20  &  0.20  &  0.20 &   0.21  \\
		     &                  &                &                    &                 &        &        &        &       &        &         &       &        &          &       &        &          &        &        &       &        \\
{\bf KK221}          &                  &                &                    &                 &        &         &        &      &        &         &       &        &          &       &        &          &        &        &       &        \\
{\bf 966}            & 2.42             & 0.67           &   2.21             & 2.25            &-0.1319 & -0.1006 &  0.61 &  1.44 &   1.05 &   0.39  & 1.79 &  -0.12 &   1.97   & 1.93  &  0.0065 &  0.0566 &  0.85  &  0.86  &  0.48 &   0.14  \\
(85) \hskip 9pt $\pm$& 0.09             & 0.09           &   0.09             & 0.10            & 0.0007 &  0.0011 &  0.04 &  0.04 &   0.05 &   0.05  & 0.06 &   0.07 &   0.07   & 0.08  &  0.0020 &  0.0020 &  0.08  &  0.08  &  0.08 &   0.09  \\
{\bf 24n}            & 3.16             & 1.46           &   3.30             & 2.43            &-0.0963 & -0.1251 &  0.88 &  1.60 &  -0.73 &   0.27  & 1.20 &   2.29 &   2.49   & 1.74  &  0.0119 &  0.0581 &  0.66  &  1.36  &  0.20 &  -0.18  \\
(35) \hskip 9pt $\pm$& 0.24             & 0.24           &   0.25             & 0.25            & 0.0019 &  0.0029 &  0.11 &  0.12 &   0.15 &   0.15  & 0.17 &   0.20 &   0.20   & 0.21  &  0.0054 &  0.0055 &  0.22  &  0.22  &  0.22 &   0.22  \\
		     &                  &                &                    &                 &        &        &        &       &        &         &       &        &          &       &        &          &        &        &       &        \\
{\bf KK084}          &                  &                &                    &                 &        &        &        &       &        &         &       &        &          &       &        &          &        &        &       &        \\
{\bf 705}            & 1.10             &-0.90           &   2.00             & 1.97            &-0.1074 & -0.0493 &  0.68 &  3.12 &   0.57 &   0.83  & 1.42 &   2.71 &   2.15   & 1.35  &  0.0333 &  0.1000 &  1.36  &  0.62  & -0.26 &   1.02  \\
(31) \hskip 9pt $\pm$& 0.21             & 0.21           &   0.22             & 0.22            & 0.0018 &  0.0027 &  0.09 &  0.10 &   0.13 &   0.13  & 0.14 &   0.16 &   0.16   & 0.18  &  0.0047 &  0.0047 &  0.18  &  0.19  &  0.19 &   0.19  \\
{\bf 830}            & 3.82             & 0.96           &   2.81             & 2.06            &-0.1075 & -0.0693 &  0.60 &  1.93 &   1.38 &   0.80  & 0.01 &  -0.58 &   1.90   & 1.83  &  0.0182 &  0.0536 &  0.80  &  1.15  &  0.35 &  -0.21  \\
(47) \hskip 9pt $\pm$& 0.07             & 0.07           &   0.07             & 0.07            & 0.0006 &  0.0009 &  0.03 &  0.03 &   0.04 &   0.04  & 0.05 &   0.06 &   0.06   & 0.06  &  0.0016 &  0.0016 &  0.06  &  0.06  &  0.07 &   0.07  \\
{\bf 666}            &-2.06             &-6.21           &   0.07             &-1.40            & 0.0343 &  0.0736 &  1.93 &  6.98 &   4.65 &   2.25  & 3.02 &   6.38 &   1.56   & 2.95  &  0.0643 &  0.1977 &  2.58  &  1.73  &  1.61 &   1.28  \\
(24) \hskip 9pt $\pm$& 0.28             & 0.29           &   0.30             & 0.30            & 0.0025 &  0.0041 &  0.13 &  0.15 &   0.18 &   0.18  & 0.19 &   0.22 &   0.22   & 0.24  &  0.0072 &  0.0073 &  0.24  &  0.25  &  0.25 &   0.25  \\
{\bf 36n}            &-1.23             &-4.00           &  -0.45             &-0.23            &-0.0374 &  0.0087 &  0.45 &  2.35 &  -0.08 &   0.62  & 2.22 &   1.16 &   1.23   & 5.91  &  0.0592 &  0.1589 &  2.60  &  2.58  &  1.97 &  -0.21  \\
(23) \hskip 9pt $\pm$& 0.19             & 0.19           &   0.20             & 0.20            & 0.0018 &  0.0026 &  0.08 &  0.09 &   0.12 &   0.12  & 0.13 &   0.15 &   0.15   & 0.16  &  0.0044 &  0.0044 &  0.17  &  0.17  &  0.17 &   0.00  \\
		     &                  &                &                    &                 &        &        &        &       &        &         &       &        &          &       &        &          &        &        &       &        \\
{\bf U3755}          &                  &                &                    &                 &        &        &        &       &        &         &       &        &          &       &        &          &        &        &       &        \\
{\bf 652}            & 3.76             &-2.35           &   3.15             & 0.83            &-0.0868 & -0.0420 &  0.73 &  2.30 &   3.12 &   0.85  & 1.85 &   1.17 &   1.83   & 1.94  &  0.0194 &  0.0665 &  1.16  &  1.02  &  0.68 &   0.24  \\
(42) \hskip 9pt $\pm$& 0.12             & 0.12           &   0.12             & 0.12            & 0.0011 &  0.0017 &  0.06 &  0.06 &   0.08 &   0.08  & 0.08 &   0.09 &   0.10   & 0.10  &  0.0028 &  0.0029 &  0.11  &  0.11  &  0.11 &   0.11  \\
{\bf 914}            & 6.72             & 4.57           &   4.47             & 4.03            &-0.1930 & -0.1193 &  0.38 &  0.35 &  -1.28 &   0.83  & 0.26 &  -1.05 &   4.31   & 0.08  & -0.0001 &  0.0661 &  1.47  &  1.28  &  0.66 &   0.36  \\
(23) \hskip 9pt $\pm$& 0.18             & 0.18           &   0.18             & 0.18            & 0.0012 &  0.0017 &  0.06 &  0.07 &   0.09 &   0.09  & 0.10 &   0.13 &   0.14   & 0.15  &  0.0036 &  0.0037 &  0.16  &  0.17  &  0.17 &   0.17  \\
{\bf 1182}           & 8.48             & 6.78           &   6.49             & 5.38            &-0.2120 & -0.1678 &  0.59 & -1.98 &  -0.41 &  -0.17  & 1.13 &  -0.34 &   5.89   & 2.17  & -0.0153 &  0.0232 &  0.25  &  0.86  & -0.30 &  -0.27  \\
(49) \hskip 9pt $\pm$& 0.08             & 0.08           &   0.08             & 0.08            & 0.0006 &  0.0009 &  0.03 &  0.04 &   0.04 &   0.05  & 0.05 &   0.06 &   0.07   & 0.07  &  0.0017 &  0.0017 &  0.07  &  0.07  &  0.08 &   0.08  \\
{\bf 1257}           & 3.26             & 1.05           &   2.84             & 2.30            &-0.0894 & -0.0769 &  0.89 &  3.08 &  -0.10 &   0.51  & 0.73 &   0.23 &   2.78   & 3.00  &  0.0197 &  0.0667 &  1.25  &  2.26  &  1.11 &   1.02  \\
(25) \hskip 9pt $\pm$& 0.20             & 0.20           &   0.20             & 0.21            & 0.0019 &  0.0026 &  0.09 &  0.10 &   0.12 &   0.12  & 0.13 &   0.16 &   0.16   & 0.17  &  0.0045 &  0.0046 &  0.18  &  0.18  &  0.18 &   0.18  \\
{\bf 2123}           & 4.94             & 2.87           &   3.41             & 3.16            &-0.1327 & -0.1067 &  0.81 & -0.72 &   1.50 &   1.31  & 2.31 &  -0.06 &   2.59   & 0.25  & -0.0020 &  0.0167 &  0.24  &  0.88  &  0.89 &   0.26  \\
(31) \hskip 9pt $\pm$& 0.14             & 0.14           &   0.14             & 0.14            & 0.0010 &  0.0014 &  0.05 &  0.06 &   0.07 &   0.07  & 0.08 &   0.10 &   0.10   & 0.11  &  0.0028 &  0.0029 &  0.12  &  0.13  &  0.13 &   0.13  \\
{\bf 2363}           & 9.32             & 6.76           &   6.03             & 6.02            &-0.1690 & -0.1133 &  0.60 & -1.07 &   0.93 &   0.31  & 1.66 &   0.04 &   5.43   & 1.20  &  0.0091 &  0.0330 &  0.47  &  0.90  &  0.66 &  -0.52  \\
(38) \hskip 9pt $\pm$& 0.10             & 0.12           &   0.11             & 0.11            & 0.0007 &  0.0012 &  0.04 &  0.05 &   0.06 &   0.06  & 0.07 &   0.08 &   0.08   & 0.09  &  0.0023 &  0.0023 &  0.10  &  0.10  &  0.10 &   0.10  \\
{\bf 2459}           & 7.58             & 7.39           &   5.88             & 5.54            &-0.2170 & -0.1784 &  0.57 & -1.55 &  -1.01 &   0.30  & 1.37 &  -0.38 &   4.92   & 1.18  &  0.0048 &  0.0442 &  1.03  &  1.27  &  0.63 &  -0.13  \\
(44) \hskip 9pt $\pm$& 0.10             & 0.10           &   0.10             & 0.10            & 0.0007 &  0.0011 &  0.04 &  0.04 &   0.05 &   0.06  & 0.06 &   0.08 &   0.08   & 0.08  &  0.0022 &  0.0022 &  0.09  &  0.10  &  0.10 &   0.10  \\
		     &                  &                &                    &                 &        &        &        &       &        &         &       &        &          &       &        &          &        &        &       &        \\
{\bf E490-17}        &                  &                &                    &                 &        &        &        &       &        &         &       &        &          &       &        &          &        &        &       &        \\
{\bf 2035}           & 4.63             & 4.94           &   4.84             & 4.94            &-0.1082 & -0.0853 &  1.18 &  0.26 &   0.21 &   0.33  & 0.11 &  -1.76 &   4.50   & 0.64  &  0.0084 &  0.0426 &  0.47  &  0.63  &  1.61 &   0.38  \\
(13) \hskip 9pt $\pm$& 0.55             & 0.56           &   0.56             & 0.57            & 0.0035 &  0.0055 &  0.20 &  0.21 &   0.26 &   0.27  & 0.30 &   0.40 &   0.41   & 0.45  &  0.0111 &  0.0113 &  0.49  &  0.51  &  0.52 &   0.53  \\
{\bf 1861}           & 2.60             & 1.10           &   2.93             & 1.90            &-0.0666 & -0.0611 &  0.32 &  1.06 &  -0.40 &   1.25  & 0.07 &  -0.38 &   1.42   & 1.17  &  0.0224 &  0.0413 &  1.10  &  0.34  &  0.16 &   0.36  \\
(17) \hskip 9pt $\pm$& 0.47             & 0.48           &   0.49             & 0.49            & 0.0038 &  0.0055 &  0.19 &  0.22 &   0.27 &   0.27  & 0.31 &   0.37 &   0.37   & 0.40  &  0.0105 &  0.0106 &  0.42  &  0.43  &  0.44 &   0.44  \\
		     &                  &                &                    &                 &        &        &        &       &        &         &       &        &          &       &        &          &        &        &       &        \\
\hline
\hline
\end{tabular}
\end{center}
\tablecomments{Approximate S/N ratios per \AA\ measured at 5000 \AA\ are listed in parentheses in the first column (second line).}
\end{table}

\clearpage


\end{document}